\documentstyle[eqsecnum,preprint,aps,epsf]{revtex}

\newcommand{\ave}[1]{\mbox{$\langle #1 \rangle$}}

\newcommand{\Eq}[1]{\mbox{Eq.~(\ref{#1})}}

\newcommand{\beq}{\begin{equation}}
\newcommand{\eeq}{\end{equation}}
\newcommand{\beqa}{\begin{eqnarray}}
\newcommand{\eeqa}{\end{eqnarray}}
\newcommand{\bmath}{\begin{mathletters}}
\newcommand{\emath}{\end{mathletters}}
\newcommand{\kstyle}{\mbox{$ {\cal K}$}}

\newcommand{\cn}{{\rm cn}}

\newcommand{\dn}{{\rm dn}}

\newcommand{\sech}{{\rm sech}}
\newcommand{\disc}{\mbox{$\delta$}}

\begin{document}
\draft
\preprint{ICTP-SISSA}
 
\title{Coherent oscillations between two weakly coupled  
Bose-Einstein Condensates: Josephson effects, $\pi$-oscillations,
and macroscopic quantum self trapping}
 
\author{S.~Raghavan$^1$, A. ~Smerzi$^2$,
S. ~Fantoni$^{1,2}$, and S. ~R. ~Shenoy$^1$}
\address{
1) International Centre for Theoretical Physics,
 I-34100, Trieste, Italy \\
2) Istituto Nazionale de Fisica della Materia and 
International School for Advanced Studies,
 via Beirut 2/4, I-34014, Trieste, Italy}
\date{\today}
\maketitle
\begin{abstract}

We discuss the coherent atomic oscillations between two weakly coupled
Bose-Einstein condensates. The weak link is provided by  
a laser barrier in a (possibly asymmetric)
double-well trap or by  Raman coupling between two condensates
in different
hyperfine levels. The  Boson Josephson Junction (BJJ) dynamics  
is described by the two-mode 
non-linear Gross-Pitaevskii equation, 
that is solved analytically in terms of elliptic functions.
The BJJ, being a $neutral$, isolated system, allows
the investigations of new dynamical regimes for the phase difference 
across the junction and for the population imbalance,
that are not accessible with
Superconductor Josephson Junctions (SJJ). These include oscillations with 
either, or both of the following properties:  
1)  the time-averaged value of the phase 
is equal to $\pi$ ($\pi-phase$ oscillations);
2) the average population imbalance is nonzero, in states with 
``macroscopic quantum self-trapping'' (MQST).
 The (non-sinusoidal) generalization of the SJJ `ac' and `plasma' 
oscillations and 
the Shapiro resonance can also be observed. 
We predict the collapse of 
experimental data (corresponding to different trap geometries
and total number of condensate atoms)
onto a single universal curve, for the inverse period of oscillations.

Analogies with  Josephson oscillations 
between two weakly coupled
reservoirs of $^3$He-B and the
internal Josephson effect in $^3$He-A 
are also discussed.

\end{abstract}
\pacs{PACS: 03.75.Fi,74.50.+r,05.30.Jp,32.80.Pj}

\section{Introduction}
\label{sec:intro}
Bose-Einstein condensation, predicted more than 70 years ago
\cite{bose-einstein}, was detected in 1995, in a weakly interacting gas
of alkali atoms held in 
magnetic traps
 \cite{bec123}. Following the first observations, there have been 
important experimental
developments. 
A superposition  of 
condensate atoms in different hyperfine levels \cite{sacimsk,smcisk} has
been created; 
non-destructive,
{\em in situ},
 detection probes have tracked
 the dynamical evolution 
of a single condensate \cite{amddkk}. Further, the evolution of the
relative phase of two condensates has been measured through 
interferometry techniques \cite{binmix98jila2}. More recently, 
experiments that tune the scattering length by several
orders of magnitude \cite{iasmsk}
have opened the definite possibility of creating
in the laboratory an ideal condensate of non-interacting atoms. 

The precise manipulation of this  form of matter
 is of considerable  theoretical 
interest: besides the study of 
 fundamental aspects of superfluidity 
from ``first principles'', it is possible to 
address  ``foundational'' problems of quantum mechanics \cite{sollegg}.
In fact, the order parameter can be identified with the one-body 
macroscopic condensate wave function. This obeys a nonlinear 
Schr\"odinger equation, known in the literature as the Gross-Pitaevskii 
equation (GPE) \cite{gpe}.
The GPE has been successfully applied to study 
kinetic properties of the condensate, like collective mode frequencies
of  trapped Bose-Einstein condensates (BEC) 
\cite{collect-edwards-stringari} and
 the relaxation times of monopolar oscillations \cite{smerfan}.
Chaotic behavior in dynamical quantum
 observables \cite{smerfan,kagan_chaos}, and the metastability of 
quantized vortices has been predicted \cite{dodd-rokh-brsfs}. 

 The existence of  spatial quantum coherence 
was demonstrated by cutting a single trapped condensate  
by a far off-resonant laser sheet. Upon switch-off of the
confining trap and the laser, the two condensates
overlapped, producing
interference fringes \cite{andrews}, in    
analogy with the a double-slit experiment with single electrons. 

However, the superfluid nature of BEC can be fully tested 
only trough the existence of superflows.
Current
experimental efforts are being focused on the
creation of a Josephson junction  
between two condensate bulks \cite{andrews,binmix98jila}. 
In this context, the  Josephson junction problem 
 has been studied  theoretically in the limit of 
non-interacting atoms \cite{javan1}, for small amplitude 
Josephson oscillations \cite{dps,zapsoleg}, 
including finite temperature (damping) effects \cite{zapsoleg}.
Furthermore,  in connection with the 
quantum measurement problem,  decoherence effects
 and  quantum corrections to the semiclassical
mean-field dynamics \cite{mcww,imalewyou} have also been studied.
Self trapping dynamics 
in the limit of small number of condensate atoms
 has been considered  \cite{mcww}
 in the ``quantum'' and in the ``semiclassical'' (mean-field)
approximation.
We have elsewhere,\cite{sfgs},
 pointed out that even though the 
Boson Josephson Junction (BJJ) 
 is
a $neutral$-atom system, it  can still display the 
(non-sinusoidal generalization of) typical `dc', `ac' and
Shapiro effects occuring in charged Cooper-pair
 superconducting junctions. Moreover, 
 novel dynamical regimes like
macroscopic quantum
self trapping (for arbitrarily large condensates)
and $\pi$-phase oscillations (where the average value of
the phase across the junction is
equal to  $\pi$), have been predicted.
 In the present paper we present 
a comprehensive  analysis of 
the effects described in \cite{sfgs}, 
including a discussion of the BJJ equations and their analytic solution, and
limits of the approximations underlying the BJJ model, and a comparison
with other superconducting and superfluid Josephson junctions. 

The description of the GPE dynamics 
for a Bose condensate in a double-well trap
reduces, under certain conditions, to a nonlinear, two-mode equation
for the time-dependent amplitudes $\psi_{1,2}(t) = 
\sqrt{N_{1,2}(t)} e^{i \theta_{1,2}(t)}$, where $N_{1,2}(t)$ and
$\theta_{1,2}(t)$ are the number of atoms and the phases of the condensate in 
the trap $1,2$ respectively. 
These amplitudes 
are coupled by a tunneling matrix element between
the two traps, with
the spatial dependence of the GPE wave function integrated out into 
constant parameters.
The resulting BJJ tunneling equations resemble the  
(nonlinear generalizations of) Superconductor Josephson Junction 
(SJJ) equations, with the variables being the relative phase and the
fractional population imbalance.

However, there are important physical differences between the
isolated double-well BJJ and 
the SJJ with an
external circuit. The SJJ  is generally discussed in terms of a 
rigid pendulum analogy in the resistively shunted junction (RSJ)
 model, while the BJJ in a double-well trap 
 can only be completely understood
in terms of a {\em non-rigid}
pendulum analogy, with a length dependent on the angular momentum.
 In SJJ the Cooper-pair population imbalance is identically
 zero (considering two equal-volume superconducting grains) due to the
presence of the external circuit\cite{barone-ohta},
and the dynamical variable is the voltage $\sim \dot{\phi}$ across
a quasiparticle resistive shunt. In the BJJ, the
non-rigid pendulum dynamics are associated with {\em superfluid density}
oscillations, of an isolated system. 
 An isolated (without external circuit) superconducting junction 
allows coherent Cooper-pair oscillations, but only 
in the small amplitude (plasma) limit \cite{barone-ohta,solymar,tinkham}. 
 
A closer analog of the BJJ is provided by the internal 
Josephson effect in $^3$He-A, where the (rigid) pendulum oscillations 
describe the rate of change of up and down spin-pair populations, induced 
by an external variable magnetic field \cite{wkw,leggrmp,matsu}. 
The ``$\pi$ oscillations'' between 
two weakly coupled reservoirs of $^3$He-B \cite{bpsldp}
 could be related to the analogous
oscillations occuring in BJJ.
                                                                 
The experimental detection of predicted effects in BJJ could be
achieved through 
temporal modulations of phase-contrast fringes \cite{andrews},
interferometric techniques \cite{binmix98jila2}, 
or other probes of atomic populations \cite{bec1},
using 
$\sim$ millisecond temporal oscillations of the
 integrated signal $N_1 - N_2$.
The direct  detection of the
 currents instead of densities, perhaps by Doppler interferometry, would
be worth exploring.
\par
The plan of the paper is as follows:
 In Section~\ref{sec:partII}, we
obtain the BJJ
tunneling equations, that, in Section~\ref{sec:partIII}, are compared with
the Josephson equation for other superconductor and superfluid systems.
In Section~\ref{sec:partIV}
 we solve the BJJ equations discussing the 
various dynamical regimes. In Appendix~\ref{sec:partX}
 we outline the 
derivation of the two-mode BJJ from GPE and discuss the limit of the
approximations. The BJJ equations are solved 
analytically in terms of elliptical functions in Appendix~\ref{sec:partXI}.
  In the remainder of the paper
Section~\ref{sec:partasym}, 
 we discuss 
the asymmetric trap case,
clarifying the analogies with the $ac$
and Shapiro effect. 
We summarize our results in Section~\ref{sec:end}.

\section{Boson Josephson junction: the non-linear two-mode approximation }
\label{sec:partII}
The  wavefunction $\Psi(r,t)$ for an interacting BEC in a trap potential
$V_{trap} (r,t)$ at $T=0$ 
satisfies the GPE:
\begin{equation}
i\hbar\frac{\partial \Psi(r,t)}{\partial t} = 
-\frac{\hbar^2}{2m}\nabla^2\Psi(r,t) +
 [V_{trap}(r) +
g_0 |\Psi(r,t)|^2]\Psi(r,t),
\label{eq:gpe}
\end{equation}
with $g_0=4\pi\hbar^2a/m$, $m$ the atomic mass, and
$a$ the $s$-wave scattering length of the atoms \cite{nota}.
In the following we will consider a double-well trap produced,
for example,
by a far off-resonant laser barrier that cuts a single trapped condensate
into two (possibly asymmetric) parts \cite{andrews}.
However,  the results could also apply to 
the oscillations of the condensate population
difference between two hyperfine levels
\cite{binmix98jila}. 

Since we are interested in the dynamical oscillations of the two weakly 
linked BEC, we write a (time-dependent) $variational$ wave-function as:
\beq
\Psi(r,t) = \psi_1(t)\Phi_1(r)+\psi_2(t)\Phi_2(r) 
\label{eq:psi12defn}
\eeq
with $\psi_{1,2}(t) = \sqrt{N_{1,2}}e^{i\theta_{1,2}(t)}$ and
a constant total number of atoms $N_1+N_2=|\psi_1|^2+|\psi_2|^2=N_T$.
The amplitudes for general occupations $N_{1,2}(t)$, and phases
$\theta_{1,2}(t)$ obey the nonlinear two-mode dynamical equations
 \cite{zapsoleg,mcww,imalewyou,sfgs,bbs,kc},
\bmath
\label{eq:gpetwost}
\beqa
i\hbar\frac{\partial\psi_1}{\partial t}&=&
(E_1^0+U_1 N_1)\psi_1-\kstyle\psi_2 \\
i\hbar\frac{\partial\psi_2}{\partial t}&=&
(E_2^0+U_2 N_2)\psi_2-\kstyle\psi_1,
\eeqa
\emath
where damping and finite temperature effects are ignored. Here
$E^0_{1,2}$ are the zero-point energies in each well, $U_{1,2}N_{1,2}$
are proportional to
the atomic self-interaction energies, and 
 $\kstyle$
describes the amplitude of the tunneling between condensates,
see Fig.~\ref{fig:schematic}. 
The constant parameters $E^0_{1,2},U_{1,2},\kstyle$ can be written in terms of 
$\Phi_{1,2}(r)$ wave-function overlaps. The $\Phi_{1,2}(r)$, 
roughly describing the condensate in each trap, can be expressed in terms
of stationary symmetric and antisymmetric eigenstates of GPE,
(see Appendix~\ref{sec:partX}).

 The fractional population imbalance
\beq
z(t)\equiv(N_1(t)-N_2(t))/N_T\equiv(|\psi_1|^2-|\psi_2|^2)/N_T
\label{eq:zdef}
\eeq
and
relative phase
\beq
\phi(t)\equiv\theta_2(t)-\theta_1(t)
\label{eq:phidef}
\eeq
obey
\bmath
\label{eq:zphidot}
\beqa
\label{eq:zphidota}
\dot{z}(t) &=& -\sqrt{1-z^2(t)} \sin(\phi(t)), \\
\label{eq:zphidotb}
\dot{\phi}(t) &=& \Delta E + \Lambda z(t) + \frac{z(t)}{\sqrt{1-z^2(t)}}
\cos (\phi(t)),
\eeqa
\emath
where we have rescaled to a dimensionless time, $t 2\kstyle/\hbar
\rightarrow t$, and
\bmath
\label{eq:delElmbdef}
\beqa
\Delta E \equiv (E_1^0-E_2^0)/2\kstyle + (U_1-U_2)/2\kstyle,\\
\Lambda \equiv U N_T/2\kstyle \; ; \; U \equiv (U_1+U_2)/2
\eeqa
\emath
The dimensionless parameters $\Lambda$ and $\Delta E$ determine the 
dynamic regimes of the BEC atomic tunneling. 
The total, conserved energy is:
\beq
H = \frac{\Lambda z^2}{2} + \Delta E z - \sqrt{1-z^2}\cos\phi.
\label{eq:ham}
\eeq
suggesting that 
the equations of motion (\ref{eq:zphidot}) can be written in
the Hamiltonian form
\beq
\dot{z} = -\frac{\partial H}{\partial \phi} 
\; \; ; \; \; 
\dot{\phi} = \frac{\partial H}{\partial z},
\label{eq:zphidotham}
\eeq
with $z$ and $\phi$, the canonically conjugate variables.

For well defined mean values in relativ population and phase, 
fluctuations must be small. 

\section{The Josephson effect in other superfluid and superconducting 
systems} 
\label{sec:partIII}
{\bf The superconducting Josephson Junction (SJJ)}.
We now consider the SJJ dynamic 
equations \cite{barone-ohta,solymar,tinkham,fulton}, for
 comparison with
the BJJ tunneling equations of Eq.~(\ref{eq:zphidot}).
The SJJ has an external closed circuit that typically includes a current
drive $I_{ext}$; the measurable developed voltage across the junction $V$ is
proportional to the rate of change of phase:
\bmath
\label{eq:SJJ}
\beqa
\label{eq:SJJ1}
I_{ext} &=& C_J \frac{dV}{dt} + I_J \sin \phi + \frac{V}{R}\\
\label{eq:SJJ2}
\dot{\phi} &=& \frac{2 e V}{\hbar}
\eeqa
\emath
where $C_J (I_J)$ is the junction capacitance (critical current) and 
$R$ is the effective resistance offered by the quasiparticle junction and
the circuit shunt resistor.  The
     $\sqrt{1-z^2}$ factors of
Eq.~(\ref{eq:zphidot}) are missing here, since the external circuit
suppresses  charge imbalances, i.e. $z(t) \equiv 0$ \cite{barone-ohta}. 
The junction charging energy $E_C \sim C_J^{-1}$;
 superconductor-grain charging energies
$E_{CG} (\sim$ inverse grain sizes) that are the analogs of interatom
interactions $U$ of the BJJ, are relevant only in mesoscopic systems.
Two such small isolated grains \cite{gold-halp} can be considered
a  closer superconducting analog of BJJ. Even in that case, as $N_T$ is
still large, the voltages that appear are 
$2 e V \sim 2 \Delta_{qp}$, the quasiparticle gap, implying
 that $|z| \sim
10^{-9}$. 
\par
Mechanical analogs have been useful in visualizing the SJJ.
Equation~(\ref{eq:SJJ}) can be written as 
\beq
\label{eq:SJJphionly}
\ddot{\phi} + \dot{\phi}/RC_J + \omega_J^2 \sin \phi =
(I_{ext}/I_J)\omega_J^2
\eeq
in unscaled units, with $\omega_J = \sqrt{E_{C} E_J}/\hbar$,
 the Josephson plasma
frequency. This can be regarded as the equation for a `particle' of `mass'
$\sim \omega_J^{-2}$ and `position' $\phi$ moving on a tilted, rigid
`washboard' potential $-\cos \phi - (I_{ext}/I_J)\phi$, with friction
coefficient $\sim 1/RC_J$. 
 Alternatively, Eq.~(\ref{eq:SJJ}) describes \cite{fulton} a rigid pendulum 
of tilt-angle $\phi$; moment of inertia $\sim \omega_J^{-2}$;
angular momentum $V \propto \dot{\phi}$ the angular velocity; damping rate
$(R C_J)^{-1}$  and external torque $\sim I_{ext}$. 

The Josephson effects in SJJ follow at once from physical
considerations: \\
{\em Plasma oscillations}: For $I_{ext} = 0$, the rigid pendulum can have
 small, harmonic oscillations in angle $\phi$ 
around the vertical.
 Linearizing Eq.~(\ref{eq:SJJ}) produces
  sinusoidal voltage/current
`plasma'  oscillations
 of  angular frequency
(in unscaled units) 
\beq
\omega \approx \omega_p \equiv 2\pi/\tau_p = \sqrt{E_c E_J}/\hbar,
\label{eq:omgsjjapprox}
\eeq
independent of initial conditions $\phi(0),\dot{\phi(0)}$. \\
 {\em ac effect}:  In the pendulum analogy,
the external drive balanced by the damping enforces steady 
rotatory motion for 
$I_{ext}/I_J > 1$. 
 The phase
increases linearly with time, $\phi(t) \sim 2eV t/\hbar$, 
where $V = I_{ext} R$ is the dc voltage developed, and the
 current oscillation
has angular frequency 
\beq
\omega = \omega_{ac} = \frac{2 \pi}{\tau_{ac}} = \frac{2eV}{\hbar},
\label{eq:acomega}
\eeq
independent of $\phi(0),\dot{\phi}(0)$. \\
{\em Shapiro resonance effect}: If a small
ac component is added to an applied dc voltage,
 $\Delta E \rightarrow \Delta E(1+\delta_0\cos\omega_0 t)$; 
$\delta_0 \ll 1$, then at resonance $\omega_0 = \omega_{ac}$, there is a
dc tunneling current with a nonzero time average
$\ave{\dot{z}(t)} \sim \delta_0\langle \sin(\omega_{ac}t+\phi(0))
\sin\omega_0 t\rangle \neq 0.$ This Shapiro resonance repeats at
higher harmonics $\omega_{ac}=2\pi/\tau_{ac}=n\omega_0,n=1,2,\ldots,$
with characteristic Bessel function coefficients $J_n(n \delta_0)$ 
\cite{barone-ohta,solymar}.
\par
Can the BJJ show the full range of SJJ effects? Not at first sight, since the
 double-well BEC is a neutral-atom system.
However,
the ability to tailor traps and the condensate
self-interaction compensates for 
electrical neutrality \cite{sfgs}.
Asymmetric positioning of the laser barrier could produce a zero-point
energy difference $\Delta E$, analogous to an applied voltage,
since the effective potential seen by the atoms on the smaller-volume
side will have a larger curvature. 
The interwell difference between the 
(bulk) nonlinear atomic self-interaction, $\sim U N_T z$ plays 
the role of a junction capacitance energy in
the dynamics.

In the SJJ, $E_J$ as well as $N_{1,2}$ are fixed 
\cite{barone-ohta,solymar}. For the
BJJ, the laser-sheet intensity, and hence the coupling $\kstyle$, can be
varied. Initial states, $N_1(0) \neq N_2(0)$,
 i.e., $z(0) \neq 0$, can be prepared,
and the laser barrier then lowered to permit tunneling.

{\bf{The internal Josephson effect in $^3$He-A}}.
A closer analog of the BJJ 
Eqs.~(\ref{eq:zphidot}) 
is provided by the longitudinal 
 magnetic resonance  in  $^3$He-A \cite{wkw}, 
that is generally understood 
as internal Josephson oscillations between two interpenetrating 
populations of superfluid up-down spin-pairs
 \cite{leggrmp}. The weak coupling 
is provided by the 
dipole interaction between pairs of up and down spins.  
The spin dynamics is governed by \cite{matsu}:
\bmath
\label{eq:hea}
\beqa
\dot{z}(t) &=& - \sin(\phi(t)), \\
\label{eq:zphidotc}
\dot{\phi}(t) &=& \Delta E + \Lambda z(t)
\eeqa
\emath
where $z(t)$ is the fractional population imbalance between
up and down spin  Cooper 
pairs,
$\Lambda \propto (\chi g_D)^{-1}$ with $\chi, g_D$ the susceptibility and
the dipole coupling respectively, and
$\Delta E \propto (B / \chi g_D)$ 
with $B$ the external applied static magnetic field.
In \cite{wkw} experiments have confirmed Eqs.~(\ref{eq:hea}), 
showing the transient between the small amplitude and 
ringing oscillations of the pendulum equations  Eqs.~(\ref{eq:hea}).

{\bf{Josephson Oscillations between two weakly linked reservoirs of $^3$He-B}}.
Quite recently the first direct experimental observation of
Josephson
oscillations between two weakly linked superfluid systems has been reported
\cite{plbdp,bpldp}.
The weak link was provided by $\sim 4000$ small holes in the rigid 
partition separating two $^3$He-B superfluid reservoirs, with the hole
diameter being comparable to the coherence length. A soft membrane
created a pressure difference across the weak link, inducing Josephson mass
current oscillations. These oscillations obey:
\bmath
\label{eq:heb}
\beqa
I(t) &=& I_c \sin(\phi(t)), \\
\label{eq:zphidotheb}
\dot{\phi}(t) &=& - {{2 m_3} \over \hbar \rho} \Delta P
\eeqa
\emath
with $2 m_3$ the mass of a $^3$He Cooper pair, $\rho$ the liquid density
and $\Delta P$ the pressure difference across the weak link being
proportional to the elastic constant of the membrane.
Small and large amplitude oscillations
 have been observed, as well as the 
driven running solutions of the phase $-\infty < \phi < \infty$ \cite{bpldp}, 
corresponding to a self-maintained population across the weak link.

By driving the soft membrane in resonance with the natural
Josephson frequency, a new metastable dynamical regime was observed,
with the time-averaged value of the phase-difference across the junction 
equal to $\pi$.  These metastable `$\pi$-oscillations'
have amplitudes
and frequencies smaller than the `stable' Josephson oscillations, into which
they decay with a life time that increases with decreasing temperature
\cite{bpsldp}. Analogous $\pi$-oscillations with similar properties
are described by BJJ (see Section~\ref{sec:partIV}). In a different context, 
$\pi$-junctions have been created with high-$T_c$ superconductors,
that reflect
the symmetry of the $d$-wave pairing state \cite{pij}.

\section{The Symmetric Trap Case, $\Delta E = 0$}
\label{sec:partIV}
{\bf Stationary Solutions.} 
For a symmetric BJJ, i.e., $\Delta E =0$, the
equations of motion Eq.~\ref{eq:zphidot} are
\bmath
\label{eq:zphidotsym}
\beqa
\label{eq:zphidotsyma}
\dot{z}(t) &=& -\sqrt{1-z^2(t)} \sin(\phi(t)), \\
\label{eq:zphidotsymb}
\dot{\phi}(t) &=&  \Lambda z(t) + \frac{z(t)}{\sqrt{1-z^2(t)}}
\cos (\phi(t)),
\eeqa
\emath
with the conserved energy:
\beq
H_0 = H(z(0),\phi(0)) = \frac{\Lambda z(0)^2}{2} - \sqrt{1-z(0)^2}\cos\phi(0).
\label{eq:sta}
\eeq

The ground state solution of the symmetric BJJ, Eq.~(\ref{eq:zphidotsym}), 
 is a symmetric
eigenfunction of the GPE with energy $E_+ = -1$ and:
\bmath
\beqa
\phi_s &=& 2 n \pi\\
\label{eq:ss1}
z_s &=& 0 
\eeqa
\emath
The next stationary state at higher energy $E_- = 1$, is an antisymmetric
eigenfuncion with:
\bmath
\beqa
\phi_s &=& (2 n + 1) \pi \\
z_s &=& 0
\label{eq:ss2} 
\eeqa
\emath
For  non-interacting atoms in a symmetric double-well potential, 
the eigenstates of the Schr\"odinger equation are always
symmetric
or antisymmetric, with $z_s = 0$.
However, because of the nonlinear interatomic 
interaction, there is a class of degenerate 
GPE eigenstates that break the $z-symmetry$:
\bmath
\label{eq:ss3}
\beqa
\phi_s &=& (2 n + 1) \pi \\
z_s &=& \pm \sqrt{1 - {1 \over \Lambda^2}} 
\eeqa
\emath
provided that $| \Lambda | > 1$. The energy for this state is 
$E_{sb} = {1\over2} \left(\Lambda+\frac{1}{\Lambda}\right)$. 

These $z$-symmetry breaking states are an artifact of the semiclassical 
limit in which the GPE has been derived. In a full quantum two mode 
approximation the eigenstates
 are always symmetric in the population imbalance:
as we will discuss later, such states have a large lifetime that
 scales exponentially with
 the total number of atoms. 

{\bf Rabi Oscillations.} 
For noninteracting atoms ($\Lambda = 0$) Eqs.~(\ref{eq:zphidotsym}) describe
sinusoidal
Rabi oscillations between the two traps with frequency 
$\omega_R = {2 \over \hbar} \kstyle$.
These oscillations are equivalent to  single atom dynamics, 
rather than a Josephson-effect arising from the interacting superfluid
condensate.  The possibility to tune the scattering length to 
values very close to zero \cite{iasmsk}, open avenues for their 
experimental observation.  

{\bf Zero-phase modes.}
These modes describe the intra-well atomic tunneling 
dynamics
with a zero time-average value of 
the phase across the junction, $\ave{\phi(t)} = 0$, and $\ave{z}=0$.  
To this dynamical class belong small and large amplitude 
condensate oscillations.  

{\em Small~ amplitude~ oscillations.}
The small amplitude, or ``plasma'' (in analogy with SJJ), 
effect follows at once from the pendulum analogy. From
Eq.~(\ref{eq:zphidotsym}),
 the BJJ is like a {\em non}-rigid pendulum of length
\beq
\label{eq:xynonrigid}
(x^2 + y^2)^{1/2} =  \sqrt{1-z^2}
\eeq
decreasing with angular momentum $z$, and with moment of inertia
$\Lambda^{-1}$.  
Linearizing Eq.~(\ref{eq:zphidotsym}), we
obtain sinusoidal oscillations with inverse
periods (in unscaled units)
\beq
\tau_L^{-1} = \sqrt{2 UN_T\kstyle + (2\kstyle)^2}/2\pi\hbar, 
\label{eq:tauptauac}
\eeq 
independent of the initial conditions $z(0),\phi(0)$. 
The comparison between Eq.~(\ref{eq:tauptauac}) and 
Eq.~(\ref{eq:omgsjjapprox}) indicates that
$2 N_T \kstyle ~(\sim N_T)$ 
is the analog of the Josephson coupling energy $E_J$, 
while $U~(\sim N_T^{-d/5}$, with $d$ the dimensionality of the system)
is the analog of the capacitive energy $E_C$. 
Since the coupling energy, fixed by the laser profile, is $\kstyle \sim A$,
the tunnel junction area
whereas the bulk interaction, $U N_T$, is independent of $A$,
  the oscillation
rate goes as 
$\tau_L^{-1}\sim A^{1/2}$.  (The plasma frequency for SJJ,
$\tau_p^{-1} \sim \sqrt{E_c E_J}$
 by 
contrast, is independent of 
$A$, as $E_J \sim A,E_c \sim A^{-1}$).

The Josephson-like length
${\mbox `}\lambda_J{\mbox '} \equiv \sqrt{\hbar^2/2m\kstyle}$, 
that governs spatial variation
along the junction, 
should be much greater than $\sim \sqrt{A}$ to justify neglect 
of
spatial
variations of $z,\phi$, i.e., to
obtain a `flat plasmon' spectrum. 
For $\kstyle = 0.1 {\rm nK}$,
one finds ${\mbox `}\lambda_J{\mbox '} \sim 10\mu {\rm m}.$ 
We will not, however, consider such spatial variations here.
The frequency of the small amplitude oscillations in BJJ are of the
order of $10 - 100$ ~Hz for typical trap parameters,
and should be compared with the frequencies 
of plasma SJJ that are of the order of GHz. 

{\em Large amplitude oscillations.}
In Fig.~\ref{fig:sym_evolve:phi00}
we display this regime of anharmonic oscillations, plotting
 $z = {{N_1 - N_2} \over N_T}$ as a function of
time, with the initial value of the phase difference $\phi(0) = 0$ and
$\Lambda=10$, 
and for increasing values of the inital population imbalance $z(0)$. 
Specifically, $z(0)$ takes on the values 0.1, 0.5, 0.59, 0.6, and 0.65 from
a) through e) respectively. 
Increasing $z(0)$ for fixed $\Lambda$, (or increasing $\Lambda$ for
fixed $z(0)$) adds higher harmonics  to the sinusoidal
oscillations,  
corresponding to large amplitude oscillations of the 
(non-rigid) pendulum.
This is shown in Fig.~\ref{fig:sym_evolve:phi00} (b,c). 
 The  period of such oscillations increases 
with $z(0)$, then decreases, undergoing  
a critical slowing down (Fig.~\ref{fig:sym_evolve:phi00}(d), dashed line),
 with a logarithmic divergence. The singularity in the
period corresponds to  the pendulum in a
vertically upright position, i.e., reaching the fixed point of 
Eq.~(\ref{eq:ss2}).

{\bf Running-phase modes: Macroscopic Quantum Self-Trapping.}
In addition to anharmonic and critically slow oscillations, other
striking effects occurin BJJ.
For instance, for a fixed value of the initial 
population imbalance, if the self-interaction parameter $\Lambda$ exceeds a
critical value $\Lambda_c$, the populations become macroscopically
self-trapped with $\ave{z} \neq 0$.
 There are different ways in which this state can be achieved
and all of them correspond to the condition (which we shall term the MQST
condition) that 
\beq
\label{eq:mqstcond}
H_0 \equiv H(z(0),\phi(0)) = {\Lambda \over 2} z(0)^2 - 
\sqrt{1-z(0)^2}\cos(\phi(0)) > 1
\eeq

In a series of experiments in which 
$\phi(0)$ and $z(0)$ are kept constant but $\Lambda$ is varied 
(by changing the geometry or 
the total number of condensate atoms, for example),
the critical parameter for MQST is 
\beq
\label{eq:lmbself-free}
\Lambda_c = \frac{1+\sqrt{1-z(0)^2}\cos(\phi(0))}{z(0)^2/2}.
\eeq
On the other hand,
changing the initial value
of the population imbalance $z(0)$ 
with a fixed trap geometry and total number of condensate 
atoms (and initial value $\phi(0))$,
$\Lambda$ remains constant and Eq.~(\ref{eq:mqstcond}) defines
a critical population imbalance $z_c$. 
As we shall see in
this and the next section, 
for $\phi(0)=0$, if $z(0)>z_c$, MQST sets in, but
for $\phi(0)=\pi$,  $z(0)<z_c$ marks the region of MQST. 
More generally, if $|\phi(0)| \leq \pi/2 $, MQST occurs for
$z(0)>z_c$ while for other values of $\phi(0)$, it occurs for 
$z(0)<z_c$.

In this section, we will discuss the type of MQST
 in which the phase difference of the order
parameter across the BJJ runs without bound; other types are discussed
later.  The phenomenon  
 can be understood through the  pendulum analogy. 
If the population imbalances are prepared such that 
the initial `angular kinetic energy' of the pendulum,
$z^2(0)$, exceeds  the
potential energy barrier height
 of the vertically displaced 
$\phi=\pi$ `pendulum orientation', there will occur a steady
self-sustained `pendulum rotation', with nonzero angular momentum
$\ave{z}$, and a closed-loop trajectory around the pendulum
support. 
For $H_0 < 1$ the population imbalance 
oscillates about a zero value. For $H_0 > 1$ the time-averaged
`angular momentum' is nonzero, $\langle z(t) \rangle \neq 0$,
with oscillations around this nonzero value
(Fig.~\ref{fig:sym_evolve:phi00}).
 MQST is a nonlinear effect
arising from 
the  self-interaction
$\sim U N_T z^2$ of the atoms. It is dependent on the trap
parameters, total number of atoms, and initial conditions, and is
self-maintained in a closed conservative system without external drives.
Although the SJJ `ac' effect in the RCSJ model involves a running-phase, 
it is clearly physically different from MQST, as it is a driven steady-state
independent of initial conditions. Moreover,
in SJJ the Cooper pair population imbalance are locked to zero by the 
external circuit.
MQST differs from single-electron  
Coulomb blockade effect that involves a single electron.
It also differs from the self-trapping of polarons \cite{kc} that
arise from single electrons interacting with a polarizable lattice:
arising, instead, from self-interaction of a {\em macroscopically large},
number of coherent atoms. 

{\bf $\pi$-phase modes}
These modes describe the tunneling 
dynamics
in which the time-averaged value of 
the phase across the junction is $\ave{\phi} = \pi$. 

The modes arise 
once more 
from the non-rigidity (momentum dependent length)
of the pendulum and 
are not observable with SJJ.  
Thei include small amplitude, large amplitude, 
and macroscopic 
self-trapped oscillations. 
The last has nonzero average population imbalanc, while $<z>=0$ 
for the others.
We summarize this behavior in the 
temporal evolution of $z(t)$ in Fig.~\ref{fig:sym_evolve:phi0pi} for 
$z(0)=0.6,\phi(0)=\pi$. $\Lambda$ takes the values 
0.1, 1.1, 1.111, 1.2, 1.25, and 1.3 in
Figs. ~\ref{fig:sym_evolve:phi0pi}(a-f)
 respectively. \\
{\em Small amplitude oscillations.}
For small $z$, Eqs.~(\ref{eq:zphidotsym}) can be linearized around the
fixed point (\ref{eq:ss2}) yielding
harmonic oscillations for $\Lambda < 1$,  with
a period (in unscaled units):
\beq
\tau_{\pi}^{-1} = \sqrt{(2\kstyle)^2-2 UN_T\kstyle}/2\pi\hbar 
\label{eq:sapi1}.
\eeq 
It is worth 
noticing
 that the ratio of the frequency of the small amplitude zero- and
$\pi$- mode phase oscillations is 
${\tau_{L} \over \tau_{\pi}} = \sqrt{{1 - \Lambda}\over {1+\Lambda}} < 1$ 
(similar to the 
$^3$He-B $\pi$-oscillations of Section~\ref{sec:partIII}).

Linearizing Eqs.~(\ref{eq:zphidotsym}) in $z$ only,
 the BJJ Eq.~(\ref{eq:zphidotsymb}) reduces to the very simple 
form:
\beq
{\ddot{\phi}} = - [\Lambda \sin(\phi) + {1 \over 2} \sin (2 \phi)] + 
O(z^2)
\label{eq:sap2}
\eeq

This suggests a mechanical analogy in which a particle of spatial 
coordinate $\phi$ moves in the potential:
\beq
V(\phi) = - \Lambda \cos(\phi) - {1 \over 4} \cos (2 \phi)  + 
O(z^2)
\label{eq:sap3}
\eeq
In Fig.~\ref{fig:phipotential} we see that $V(\phi)$ has a small 
valley  around $\phi = \pi$ where the
particle can oscillate. The depth of this valley decreases as
$\Lambda \to 1$. The valley persists, in the full
potential for $V(\phi)$, retaining all the higher order terms in $z$.

{\em Large amplitude oscillations}
For $\pi$-phase oscillations, the momentum dependent length allows 
the pendulum bob
to make inverted 
anharmonic oscillations with $\ave{z}=0$ 
around the (top of the) vertical axis.
For large amplitude $z(t)$ oscillations, $\Lambda$ can exceed unity, as
shown in Fig.~\ref{fig:sym_evolve:phi0pi}(b).

{\em oscillations with macroscopic quantum self-trapping}
Here teh non-rigidity allows the penulum bob to make a closed 
$<z> \ne 0$ rotation loop around the top of the vertical axis. 
There are two kinds of such $\pi$-phase modes with MQST:
those where the time average $\ave{z} < |z_s| \neq 0$, and those 
where $\ave{z} > |z_s| \neq 0$, with
 $z_s$ being the
stationary $z$-symmetry breaking value of the GPE. These two kinds
of MQST
 are shown
in the time evolution of $z(t)$ in Figs.~\ref{fig:sym_evolve:phi0pi}(d-f). 
In Fig.~\ref{fig:sym_evolve:phi0pi}(d), the system is in 
the first type of trapped state. A change over 
occurs at 
the stationary state
(Fig.~\ref{fig:sym_evolve:phi0pi}(e),dashed line).
 Once $\Lambda$ exceeds this
value $\Lambda_s=1/\sqrt{1-z(0)^2}$ (cf. Eq.~\ref{eq:ss3}),
 the system goes into 
the second type of $\pi$-phase trapped state
 (Fig.~\ref{fig:sym_evolve:phi0pi}(f)). 
\par
In order to see these different kinds of (running- and $\pi$-phase)
MQST modes more transparently, one can use the energy $H=H_0$ of
Eq.~(\ref{eq:sta}) to
write the system of equations (\ref{eq:zphidotsym}) in terms of an equation
of motion of a classical 
particle whose coordinate is $z$, moving in a potential
$W(z)$ with total energy $W_0$,
\bmath
\label{eq:zpotl}
\beq
\label{eq:zpotla}
\dot{z}(t)^2 + W(z) = W_0,
\eeq
where
\beq
\label{eq:zpotlb}
W(z) = z^2\left(1-\Lambda H_0 + \frac{\Lambda^2}{4}z^2 \right) \; ; \;
W_0 = W(z(0)) + \dot{z}(0)^2
\eeq
\emath
Figure~\ref{fig:zpotential}
 displays the potential $W(z)$ against $z$
(Figs.~\ref{fig:zpotential}(a) and (c))
and the
corresponding evolution of $\phi(t)$ (Figs.~\ref{fig:zpotential} (b) and (d))
to display the various dynamical regimes. 
In Fig.~\ref{fig:zpotential}(a) and (b) $\phi(0)=0,\Lambda=10$
 and in Fig.~\ref{fig:zpotential} (c) and (d)
$\phi(0)=\pi,\Lambda=2.5$. 
The horizontal lines indicate the energy value $W_0$.
For fixed value of $\Lambda$, and $\phi(0)=0$,
increasing the value of $z(0)$ changes $W(z)$
 from a parabolic to
a double-well. The motion of the particles lies within the classical
turning points in wich the total energy equals the potential energy.
For $z(0)=0.1$, in Fig.~\ref{fig:zpotential}(a),
 the potential is parabolic, and the 
(small amplitude) oscillations are sinusoidal. For $z(0)=0.6$
the trajectory of $z(t)$ becomes markedly nonsinusoidal, given the double-well
structure of  $W(z)$.
 For $z(0) \ge 0.6$ the total energy is
smaller than the potential barrier,
 forcing the particle to become localized in one of the 
two wells. The symmetry of the classical motion is broken.
 This corresponds to an
 MQST state.
 Figure~\ref{fig:zpotential}(b) displays the corresponding phase
  $\phi = \arccos [(\Lambda z^2/2 -H_0)/\sqrt{1 - z^2}]$ versus $z$.
  For untrapped oscillations, the $\phi,z$ trajectory is a closed 
curve, with a time-averaged value of $\phi(t) = 0$.
 In the running-mode MQST
regime $- \infty < \phi(t) < \infty$.
for the corresponding $\phi$ evolution.
\par
Let us now focus our attention on
 Figs.~\ref{fig:zpotential}(c,d).
 For $\Lambda=2.5,\phi(0)=\pi$, the
$z$-potential always has a double-well structure and the system is
self-trapped for all values of $z(0)$. For small values
of $z(0)$, the phase $\phi(t)$ is unbounded and the system exhibits
running-phase MQST. However, above a certain value of
 $z(0) = 2z_s = 2\sqrt{1-1/\Lambda^2}$, the phase
$\phi(t)$ becomes localized around $\pi$ and remains bounded for all
larger values of $z(0)$. In Fig.~\ref{fig:zpotential}(c,d),
 $z(0)=0.7$ and $0.98$ mark 
the two
 different kinds of $\pi$-phase MQST since they are on either side of
the stationary state value of $z_s = \sqrt{1-1/\Lambda^2}$. This 
point will
become more 
clear in the phase-plane portrait of Fig.~\ref{fig:phaseplane}. 

{\bf Discussion of Results.} 
A clear observational feature of the behavior of the system is the
time-period of oscillations. To this end, we  plot in
Fig.~\ref{fig:symperiodave}, 
the inverse period $1/\tau$ as a function of the ratio between the initial
population imbalance $z(0)$ and the critical population imbalance $z_c$.
Figure~\ref{fig:symperiodave}(a)
 shows the case for $\phi(0)=0$ and $\Lambda = 10 (z_c=0.6)$
(dashed line) and $\Lambda = 100 (z_c=0.2)$ (solid line). 
The initial parts of the graph for $z(0) \ll z_c$ mark sinusoidal small
amplitude (`plasma')
oscillations (Fig.~\ref{fig:sym_evolve:phi00}(a)).
On increasing  $z(0)$, the
oscillations become highly anharmonic, with the
inverse period that first 
increases, then decreases, displaying a critical
slowing down.
 The logarithmic divergence of the
 time-period at $z(0)=z_c$ is marked by the
hyperbolic secant evolution of $z(t)$
 (Fig.~\ref{fig:sym_evolve:phi00}(d)). 
 In the inset, we show the
average value, $\ave{z}$, as a function of $z(0)/z_c$. MQST is signaled by
the sharp (`phase-transition'-like) rise of $\ave{z}$ from zero to a
non-zero value. 
 For $\phi(0)=\pi$, Fig.~\ref{fig:symperiodave}(b),
something different happens. MQST occurs for values of the initial
imbalance $z(0)$ less than $z_c$. At $z(0)=z_c$ the time-period
diverges, and for larger values of $z(0)$, MQST disappears. 
\par
The dynamical behavior of the BJJ system can be summarized quite
conveniently in terms of a phase portrait of the two dynamical variables $z$
and $\phi$, as shown in Fig.~\ref{fig:phaseplane}.
 The trajectories are calculated for
different values of $\Lambda/\Lambda_c$ with $z(0)$ kept constant at 0.6.
 The light solid lines mark the evolution for
the evolution where the phase $\phi$ oscillates around 
0 and $\ave{z}=0$. The running mode MQST is shown by the 
trajectories with small dots 
for $\Lambda/\Lambda_c=1,1.5$ with the initial condition being
$\phi(0)=0$. 
 Note that for a rigid pendulum (without the
$\sqrt{1-z^2}$ term in the Hamiltonian in Eq.~(\ref{eq:ham})), one would
obtain only the curves described thus far.
 However, for the BJJ, due to
the momentum dependent potential in Eq.~(\ref{eq:ham}), there is
considerable richness as 
exhibited by the dark solid lines, dashed lines, and
 lines with large dots. All these curves correspond to $\phi(0)=\pi$.
 Note, for instance,
that as $\Lambda/\Lambda_c$ increases and approaches unity,
 the area enclosed by the
trajectory shrinks and is pinched at $\Lambda=\Lambda_c$ marking the onset
of $\pi$-phase MQST with $<z> < |z_s|$ (dashed line).
 Upon further increase, the area collapses to a point
at the $z = z_s$ stationary $z$-symmetry breaking state. Further increase of
$\Lambda/\Lambda_c$
 induces a reflection of the trajectory about the fixed point,
and $\pi$-phase MQST with $<z> > |z_s|$
(lines with large dots).
 Finally, the trajectories join the running mode MQST for
$\Lambda/\Lambda_c=2.7$ (lines with small dots). 
\par
We now outline a possible procedure for a) bridging experimental
data with our theoretical model, and b) collapsing data from different
experiments onto a single universal curve.
We note, at the very outset, that other procedures could
be experimentally
more accessible, particularly since novel methods of tailoring traps
\cite{sacimsk}
and the possibility of tuning the scattering length of atoms
\cite{iasmsk} have become current.
The calculation of the  values of $\Lambda$ and $\kstyle$
from the experimental data (for a given trap geometry and total number of 
condensate atoms) is straightforward. The onset of MQST
is provided by Eq.~(\ref{eq:lmbself-free}),
 which immediately gives the value of $\Lambda_c$
from the (experimentally imposed) initial conditions $z(0)$ and $\phi(0)$. 
Moreover, in the small amplitude limit the inverse period of the oscillations,
given by Eq.(\ref{eq:tauptauac}),
 provides the value of $\kstyle$ from the previously 
calculated $\Lambda$.
  
Different experiments done by varying arbitrarily 
the trap geometry and the number of condensate atoms
give a set of parameters $\Lambda, \kstyle$. The data collapse onto a 
single universal curve of $\pi/(\kstyle C \Lambda^2 \tau)$
versus $k^2(\Lambda)$ of Eq.~(\ref{eq:tau=}), as shown in
Fig.~\ref{fig:symperiod}.

The parameters $U N_T$, and $E^0$ can be estimated 
to be
$\sim 100 {\rm nK}$ and $\sim 10 {\rm nK}$ respectively 
for $N_T=10^4$ if we take the  trap-frequency $\omega_{trap}$
to be $\sim 100 {\rm Hz}$. $\Lambda=UN_T/2\kstyle$  can be varied
widely by changing $N_T$ or the barrier height $\sim \kstyle$ that 
depends exponentially on the laser-sheet thickness. 
Typical frequencies are then 
$1/\tau_L \sim 100 {\rm Hz}$.
With collective mode excitation energies $\Delta_{coll}\sim E^0$, and
quasiparticle gaps $\Delta_{qp} \sim \sqrt{U N_T E^0}$, 
for $U N_T z < \Delta_{qp,coll}$ intra-well excitations are not induced. 
At nonzero temperatures, BEC depletion and thermal fluctuations
will renormalize the parameters in 
Eq.~(\ref{eq:delElmbdef}),  and will damp \cite{zapsoleg}
the coherent oscillations. The effects of damping on the oscillation 
behaviour requires a separete treatment and will be considered elsewhere. 

\section{The Asymmetric Trap Case, $\Delta E\neq 0$}
\label{sec:partasym}
{\bf Exact solutions and temporal behavior.}
Let us now consider the case where the traps are asymmetric,
 i.e., $\Delta E \neq 0$, as in Fig.~\ref{fig:schematic},
with Hamiltonian
\begin{equation} \label{eq:asymham}
H = \frac{\Lambda z^2}{2} + \Delta E z - \sqrt{1-z^2}\cos\phi.
\end{equation}
For $\Lambda z(0) \ll \Delta E$, the non-rigid pendulum is driven to rotate
in a direction determined by $\Delta E$ (corresponding to the ac
Josephson-like effect). With $\Delta E=0$, and $\Lambda > \Lambda_c$ (of
Eq.~(\ref{eq:lmbself-free}), we had
found that the pendulum also executes rotatory motion, in a direction
determined by $z(0)$. For $\Lambda z(0) \gg \Delta E \neq 0$, we expect this
type of motion to persist (corresponding to MQST due to nonlinearity). In
between there should be competition between the two effects, and a
transition at some shifted critical value $\Lambda = \Lambda_c(\Delta E)$. 
This physical picture for $\Delta E \neq 0$ is confirmed by obtaining 
$z(t)$ in terms of Weierstrassian elliptic functions that change their
behavior at a singular value $\Lambda = \Lambda_c(\Delta E)$. 

We  show in Figs.~(\ref{fig:asym_period}) 
that the MQST phenomena, (inverse-period dip, average non-zero
imbalance), persist in the $\Delta E \neq 0$ case, and display dependence on
$\Lambda$ and $\Delta E$. 
Figure~\ref{fig:asym_period} shows the scaled inverse
 period $\tau_{ac}/\tau$ versus  the 
scaled nonlinearity
ratio $\Lambda/\Lambda_c(\Delta E)$ 
 where $\tau_{ac}$ is as in Eq.~(\ref{eq:tauptauac}),
with $z(0)=0.1,\Delta E=1.0$, and $\phi(0)=0,\pi$. 
The dip to zero at the onset of MQST is clearly seen. The inset shows
the time-averaged $\ave{z}$ for $\phi(0)=0,\pi$, vanishing at
$\Lambda=\Lambda_c(\Delta E)$. 
Whereas for $\Delta E = 0$ and $\Lambda < \Lambda_c(\Delta E=0)$, the
average population imbalance was zero, for $\Delta E \neq 0$ we have
$\ave{z} \neq 0$ in the corresponding sub-critical region $\Lambda <
\Lambda_c(\Delta E)$. 
This is analogous to a voltage 
across a capacitor inducing a charge difference and the
external static magnetic field in the case of $^3$He-A
 Note that there
is a combined influence of $\Lambda,\Delta E,\phi(0)$, so $\ave{z}$ can
be larger (in magnitude) than $z(0)$. In particular, for $\Lambda
\rightarrow 0$, $\ave{z} \rightarrow -\Delta
E(\sqrt{1-z^2(0)}\cos\phi(0)-\Delta E z(0))/(1+\Delta E^2)$, 
that for $\phi(0)=0$ is negative, as in the inset. This corresponds to an
averaged pendulum rotation $\ave{z}\sim-\Delta E < 0$, opposite in sign
to the initial $z(0)>0$, but slowing to zero as the critical value is
approached. For $\Lambda > \Lambda_c(\Delta E)$, in the MQST regime, the
averaged rotation $\ave{z}>0$ is in the initial direction of $z(0)>0$, with
$\ave{z}$ approaching the initial $z(0)$ value for large $\Lambda$, as in 
the $\Delta E=0$ case of Fig.~\ref{fig:symperiod}.

{\bf Shapiro Effect Analogs.}
Let us now consider the BJJ analog of the Shapiro resonance effect observed
in SJJ \cite{solymar}. In addition to a time-independent
trap asymmetry $\Delta E$, we impose a sinusoidal variation
so that we can write the asymmetry term as
 $\Delta E + \Delta E_1 \cos \omega_0 t$.
This could be done by varying the laser barrier position at
fixed intensity. A similar Shapiro-like resonance effect could be seen,
with an oscillation of the laser beam intensity, at fixed mid-position,
so $\kstyle \rightarrow \kstyle(1+\delta_0 \cos \omega_0 t)$.
The analog of the Shapiro effect
arises when the period from the time independent asymmetry,
$\sim 1/\Delta E$,
matches that from the oscillatory increment,
$ \sim 1/\omega_0$.  This
matching condition is intimately connected  with the phenomenon of 
Bloch oscillations and dynamic localization in crystals
and trapping in two-level atoms \cite{rkdbs-note}.
The
dc value of the drift 
current,  $\langle\dot{z}(t)\rangle$,
 as a function of $\Delta E$, will show up
as resonant spikes.
(For SJJ, with current drives, the Shapiro effect shows up as 
steps in the I-V characteristics.)
Of course, the dc drift cannot persist indefinitely, because
 the phase difference between the condensates on the two parts of the BJJ 
will cease to be a well-defined quantity once the population in one well 
drops below $N_{min}$. 
\par 
Figure \ref{fig:shapiro} shows $I_{dc} \propto \langle\dot{z}(t)\rangle$
obtained from time averaging the numerical solution, with a small ac drive
and $\Delta E \neq 0$. It is plotted
as a function of $\Delta E/\omega_0$ for increasing values of the
nonlinearity ratio $\Lambda$. The initial conditions are
$z(0)\sim 0 = 0.045,\phi(0)=\pi/2$, for which $\Lambda_c  \sim 1000$
(in the  absence of $\Delta E$ and ac driving). 
 When $\Lambda$ is zero, 
sharp peaks in $I_{dc}$ occur at the usual `Shapiro' condition values
 $\Delta E \propto n\omega_0, n=1,2,\ldots$.
 As $\Lambda$ increases, however, two things happen.
Firstly, multiple peaks also occur at
$\Delta E/\omega_0$
 values different from integers. 
Close to the MQST regime, ($\Lambda \sim \Lambda_c$), there is
a proliferation of peaks as the system moves from a 
regime of constant current, 
$\langle \dot{z} \rangle \neq 0$ ($\Lambda$ small), to one of 
constant population imbalance
 $\langle z \rangle \neq 0$ ($\Lambda$ large). 
Secondly, the  magnitude of the peaks or dc currents 
decreases. 
\par
Finally, we note that for $\Delta E$ larger than the Bogoliubov
quasiparticle gap $\Delta_{qp}$, and high enough temperatures, a dissipative
quasiparticle branch might be observable. 
\section{Summary}
\label{sec:end}

We have investigated the Josephson dynamics
in two weakly linked Bose-Einstein condensates forming a Boson Josephson
junction. In the resulting nonlinear two-mode model,
we have described the temporal oscillations of the population imbalance of
the condensates in terms of elliptic functions.
Our predictions include
non-sinusoidal generalizations of Josephson `dc', `ac' and Shapiro
 effects. We also predict macroscopic quantum self-trapping which is
 a self-maintained population imbalance 
across the junction due to  atomic self-interaction,   and
$\pi$-oscillations, in which the phase difference across the junction
oscillates around $\pi$. We clarify the connection and the differences 
between these phenomena
and others occuring in related systems like the superconducting
Josephson junctions, the internal Josephson effect in $^3$He-A, and
Josephson oscillations between two weakly linked reservoirs of $^3$He-B.
Through a set of functional relations, we also predict the 
 collapse of experimental data (corresponding to different trap geometries
and total number of condensate atoms)
onto a single universal curve.
These effects constitute
experimentally testable signatures of quantum phase coherence and
the superfluid character of weakly interacting Bose-Einstein
condensates.
\par
Discussions with V.~Chandrasekhar, S.~Giovanazzi,
L.~Glazman, A.~J.~Leggett, E.~Tosatti and
useful references from
G.~Williams are acknowledged.

\appendix
\section{Microscopic derivation of the Boson Josephson equation from the 
Gross-Pitaevskii equation}
\label{sec:partX}
The values of the constant parameters in BJJ Eqs.~(\ref{eq:gpetwost})
$\kstyle$, $E_0$, and $U$ depend on the 
geometry (and effective dimensionality) of the system 
and the total number of condensate atoms. We now outline their 
dependence in term of spatial GPE wave functions, 
elucidating the approximations underlying the BJJ equations.

We look for the solution of the (time-dependent) GPE~(\ref{eq:gpe}) with the
following $variational$ ansatz:
\beq
\Psi(r,t) = \psi_1(t)\Phi_1(r)+\psi_2(t)\Phi_2(r)
\label{eq:va}
\eeq
There are two approximations underlying this ansatz: 
\renewcommand{\theenumi}{\arabic{enumi}}
\begin{enumerate}
\item  We describe the temporal evolution of the
GP wave function as
the {\em superposition} of two wave functions (roughly) describing 
the condensate in each trap. The nonlinear interaction in GPE destroys
such superposition. In effect if 
the condensate density in the tunneling region is small (as it is the case 
for weak links) 
nonlinear interaction in that region is negligible, and the superposition
ansatz is preserved.
\item We factorize  the temporal and the spatial dependence 
of the GPE wave function describing the condensate in each trap. 
Later in this section,
 we will discuss the limit of validity of this approximation.  
\end{enumerate}
The spatial dependence of $\Phi_{1,2}(r)$ can be constructed by the exact
symmetric $\Phi_+(r)$ and antisymmetric $\Phi_-(r)$ stationary 
eigenstates of the GPE (see Section~\ref{sec:partIV}):
\bmath
\label{eq:va2}
\beqa
\Phi_1(r) &=& {{\Phi_+ + \Phi_-}\over 2} \\
\Phi_2(r) &=& {{\Phi_+ - \Phi_-}\over 2}
\eeqa
\emath
ensuring that:
\beq
\int{\Phi_1(r) \Phi_2(r) dr} = 0
\label{eq:va3}
\eeq
and where we impose the normalization condition:
\beq
\int{|\Phi_{1,2}(r)|^2 dr} = 1
\label{eq:va4}
\eeq
Replacing Eqs.~(\ref{eq:va},\ref{eq:va2})
 in the GPE (\ref{eq:gpe}), and using the orthogonality condition 
Eq.~(\ref{eq:va3}), we obtain the BJJ equations,
\bmath
\beqa
i\hbar\frac{\partial\psi_1}{\partial t}&=&
(E_1^0+U_1 N_1)\psi_1-\kstyle\psi_2 \\
i\hbar\frac{\partial\psi_2}{\partial t}&=&
(E_2^0+U_2 N_2)\psi_2-\kstyle\psi_1,
\eeqa
\label{eq:va5}
\emath
with constant parameters :
\bmath
\beqa
E_0 &=& \int{{\hbar^2 \over {2 m}} |\nabla \Psi |^2 +
 |\Psi|^2 V_{ext}(r) dr} \\
U &=& g_0 \int{|\Psi|^4 dr} \\
\kstyle &\simeq& - \int{ \left[
 {\hbar^2 \over {2 m}} (\nabla \Psi_1 \nabla \Psi_2 )
+\Psi_1 V_{ext} \Psi_2 \right]  dr}
\label{eq:va6}
\eeqa
\emath
We now return to our variational ansatz
 $\Psi(r,t) =\psi_1(t)\Phi_1(r)+\psi_2(t)\Phi_2(r)$.
The parameters 
 $U,\Delta E \sim$ wave-function
overlaps are $N_T$-dependent, but are independent of $z(t)$, so the
chemical potential difference is considered linear in $z$. This approximation
captures the dominant $z$-dependence of the tunneling equations coming from
the scale factors $\psi_{1,2}\propto\sqrt{N_{1,2}}$, but ignores shape changes
in the wavefunctions for $N_1(t) \neq N_2(t)$.
We can estimate such corrections to the chemical
potential difference $\Delta \mu \equiv \mu_1 - \mu_2$, within the 
Thomas-Fermi approximation $\mu_{1,2} \sim N_{1,2}^{2/5} \sim
(N_T/2)^{2/5}(1\pm z)^{2/5}$. Then relative corrections to the linear form
$\Delta \mu = (4/5) z$ are estimated by ${\cal E} \equiv (\Delta \mu(z) -
4z/5)/\Delta \mu(z)$ where $\Delta \mu(z) = (1+z)^{2/5}-(1-z)^{2/5}$. We
find that ${\cal E}$ is negligible over the $z$ range where MQST effects are
expected: ${\cal E} \sim 0.1\%$ for $z=0.1$ and ${\cal E}\sim 3\%$ for
$z=0.4$. Thus Eq.~(\ref{eq:zphidot}) with $\Delta E,\Lambda$ treated 
as constants,
is indeed a reliable nonlinear equation describing BJJ dynamics
for a large range of $z(t)$ values. Similar conclusions has been 
reached in \cite{zapsoleg}.
As a further test, the GPE~(\ref{eq:gpe})
 has been solved numerically (in a spatial grid)
in the double well geometry \cite{giovan}, fully confirming the conclusions
just outlined.

\section{Exact solutions in terms of Jacobian elliptic functions}
\label{sec:partXI}
The total energy of the system is given
by
\beq
H(z(t),\phi(t)) = \frac{\Lambda z^2}{2} + \Delta E z - \sqrt{1-z^2}\cos\phi
= H(z(0),\phi(0)) \equiv H_0,
\label{eq:basicHam}
\eeq
where $H_0$ is the initial (and conserved) energy. 
Combining Eqs.~(\ref{eq:zphidota}),(\ref{eq:ham}),
we have
\beq
\dot{z}^2 + \left[\frac{\Lambda z^2}{2} + \Delta E z - H_0\right]^2
= 1 - z^2.
\label{eq:zzdot}
\eeq
The nonlinear GPE tunneling equations for the macroscopic amplitudes
$\psi_1(t),\psi_2(t)$ are formally identical to equations governing
a physically very different problem --- a single electron in a 
polarizable medium, forming a polaron \cite{kc}.
 Solutions have been found
\cite{kc,samos-ecu,scatter,interplay},
 for the discrete nonlinear Schr\"odinger equation (DNLSE) 
describing the motion of the polaron between two sites of a dimer.
 Similarly, we use
\Eq{eq:zzdot} to obtain the exact solution for $z(t)$ in
terms of quadratures,
\beq
\frac{\Lambda t}{2} = \int_{z(t)}^{z(0)} {\frac{dz}
{\sqrt{\left(\frac{2}{\Lambda}\right)^2(1-z^2) - 
\left[z^2 + \frac{2z\Delta E }{\Lambda} - \frac{2H_0}{\Lambda}\right]^2}}}
\label{eq:basicexact}
\eeq
We consider $\Delta E=0$, and $\Delta E\neq0$ cases separately.
 For symmetric double wells,
 $\Delta E = 0$,  the denominator of Eq.~(\ref{eq:basicexact}) can be 
factorized, so
\beq
\frac{\Lambda t}{2} = \int_{z(t)}^{z(0)} {\frac{dz}
{\sqrt{(\alpha^2+z^2)(C^2-z^2)}}},
\label{eq:tzsymmetric}
\eeq
where
\bmath
\label{eq:cazdef}
\beqa
C^2 &=& \frac{2}{\Lambda^2}\left[(H_0\Lambda-1)+\frac{\zeta^2}{2}\right] \;
 ; \; 
\alpha^2 = \frac{2}{\Lambda^2}\left[\zeta^2-(H_0\Lambda-1)\right] ,\\
\zeta^2 (\Lambda) &=& 2\sqrt{\Lambda^2+1-2 H_0\Lambda}.
\label{eq:calphazetadef}
\eeqa
\emath
 The solution 
to Eq.~(\ref{eq:tzsymmetric}) is written in 
terms of the `cn' and `dn' Jacobian elliptic functions, (with 
$k$, the elliptic modulus \cite{byrd-milne})as
\bmath
\label{eq:zoftsymfull}
\beqa
\label{eq:dnlsesol}
z(t) &=& C \cn[(C\Lambda/k)(t-t_0),k] \; {\rm for}\; 0<k<1 \nonumber \\
     &=& C \dn[(C\Lambda)(t-t_0),1/k] \; {\rm for}\; k>1; \\
\label{ksq=}
k^2 &=& \frac{1}{2}\left(\frac{C \Lambda}{\zeta(\Lambda)}\right)^2 = 
\frac{1}{2}\left[1 + \frac{(H_0\Lambda-1)}{\sqrt{\Lambda^2+1-2 H_0\Lambda}}
\right], \\
\label{t0=}
t_0 &=& 2[\Lambda\sqrt{C^2+\alpha^2}F(\arccos(z(0)/C),k)]^{-1},
\eeqa
\emath
where  $F(\phi,k)=\int_0^\phi d\phi(1-k^2\sin^2\phi)^{-1/2}$
is the incomplete elliptic integral of the first kind. 
\par
The Jacobian elliptic functions 
$\cn(u,k)$ and $\dn(u,k)$ are periodic in the argument $u$ with 
period $4 K(k)$ and $2 K(k)$ respectively 
 where $K(k) \equiv F(\pi/2,k)$ is the complete elliptic integral
of the first kind. 
 The character of the solution changes when
 elliptic modulus $k=1$. 
From Eq.~(\ref{ksq=}), this mathematical
condition or singular parameter-dependence of
the elliptic functions
corresponds to the physical condition $H_0=1,\Lambda=\Lambda_c$ 
of Eq.~(\ref{eq:lmbself-free}), for the
onset of MQST: $k^2(\Lambda_c)=1$.
When $k^2 \ll 1$, $\cn(u,k) \approx \cos u + k^2 \sin u
(u-\frac{1}{2}\sin2u)$ is almost sinusoidal. When $k^2$ increases, the
departure from simple sinusoidal forms becomes drastic. For $k^2 \lesssim 1$, 
$\cn(u,k) \approx \sech u -\frac{1-k^2}{4}(\tanh u\sech u)(\sinh u\cosh
u-u)$ becomes non-periodic. When $k^2 \gg 1$, the behavior is again
periodic (but about a non-zero average):
 $\dn (u,1/k) \approx 1 - (\sin^2 u)/2 k^2$.
\par
The  time-period
of oscillation of $z(t)$ is given \cite{byrd-milne} by
\bmath
\label{eq:tau=}
\beqa
\tau &=& \frac{4 k K(k)}{C \Lambda} \; \; {\rm for}\; 0<k<1,\\
     &=& \frac{2 K(1/k)}{C \Lambda} \; \; {\rm for}\; k>1.
\eeqa
\emath
In the linear limit, $ \tau \rightarrow
\pi/\sqrt{1+\Lambda}$ in agreement with the expression for $\tau_p$ in
Eq.~(\ref{eq:tauptauac}). 
As $k\rightarrow 1$, or $\Lambda \rightarrow \Lambda_c$, the
 period becomes infinite, as in `critical slowing down', diverging
logarithmically, $K(k)\rightarrow \log(4/\sqrt{1-k^2})$. 
The evolution of the imbalance is given, in this special case, by
the non-oscillatory hyperbolic secant
($C=2\sqrt{\Lambda_c-1}/\Lambda_c$): 
\beq
z(t) = C \cn[(C\Lambda_c)(t-t_0),1] = 
C \sech C\Lambda_c(t-t_0)\;\;
{\rm for}\; k=1.
\label{eq:sechevolve}
\eeq
We now turn to the $\Delta E\neq 0$ case. 
The general form of the integral of Eq.~(\ref{eq:basicexact}) is split into
two parts,
\begin{equation} \label{eq:split}
\frac{\Lambda t}{2} = \frac{\Lambda t_0}{2} + \int_{z(0)}^{z_1}
\frac{dz'}{\sqrt{f(z')}} 
\end{equation}
where $\Lambda t_0/2$ is the integral from $z_1 $ to $z(0)$, and $z_1$
is a root of the quartic $f(z)=
\left(\frac{2}{\Lambda}\right)^2(1-z^2) - 
\left[z^2 + \frac{2z\Delta E }{\Lambda} - \frac{2H_0}{\Lambda}\right]^2$. 
Taylor expanding $f(z)$ around $z$, and with the change of variable 
$y=y(z)=(f'(z_1)/4)(z-z_1)^{-1}+f''(z_1)/24$ for which $y(z_1)=\infty$,
the integral in Eq.~(\ref{eq:split}) is cast in a standard form
\begin{equation} \label{eq:weir-integ}
\frac{\Lambda (t-t_0)}{2} =
 \int_y^\infty \frac{dy'}{\sqrt{4y'^3-g_2 y' -g_3}},
\end{equation}
which can be inverted as a Weierstrassian elliptic function
$y = \wp(\Lambda (t-t_0)/2;g_2,g_3)$. Thus 
\begin{equation} \label{eq:asymsoln}
z(t) = z_1 + \frac{ f'(z_1)/4}{ \wp(\Lambda(t-t_0)/2;g_2,g_3) - 
f''(z_1)/24}
\end{equation}
In Eq.~(\ref{eq:weir-integ}), the constants in the
cubic $h(y)=4y^3-g_2 y - g_3$ are determined from the coefficients 
${a_i}$ of $f(z)=\sum_{l=0}^4 a_{l+4} z^l$ as 
\begin{equation} \label{eq:g2g3}
g_2 = -a_4 -4a_1a_3 + 3a_2^2\;;\; 
g_3 = -a_2 a_4 + 2 a_1 a_2 a_3 - a_2^3 + a_3^2 - a_1^2 a_4,
\end{equation} 
where
\begin{equation} \label{eq:a1234}
a_1 = -\frac{\Delta E}{\Lambda}\;;\;
a_2 = \frac{2}{3\Lambda^2}(\Lambda(H_0+1)-\Delta E^2)\;;\;
a_3 = \frac{2H_0\Delta E}{\Lambda^2}\;;\;
a_4 = \frac{4(1-H_0^2)}{\Lambda^2},
\end{equation}
The solution Eq.~(\ref{eq:asymsoln}) is equivalent to that found in
polaronic \cite{tsiphd} and other contexts \cite{nmer,kenkusmodel},
where $\Delta E \neq 0$ corresponds to a difference or `disorder' in
on-site electronic or excitonic energies. 
\par
For $\Delta E = 0$ we found that the elliptic modulus $k^2$ governed 
the behavior of the Jacobian elliptic function solutions. For $\Delta E \neq
0$, the discriminant 
 $\disc = g_2^3 - 27 g_3^2.$ of the cubic $h(y)$ (with roots $y_{1,2,3}$)
governs the behavior of the Weierstrassian elliptic functions \cite{byrd-milne} 
For $\disc \neq 0$, the solutions are oscillatory about a non-zero average,
$\ave{z} \neq 0$. For $\disc=0,\ave{z}=0$, and the time-period diverges,
corresponding to $\Lambda=\Lambda_c(\Delta E)$, the onset of MQST. 
\par
The time-period of oscillation can be written in terms of complete elliptic
integrals of the first kind $K(k)$ as in the $\Delta E=0$ case of
 Eq.~(\ref{eq:tau=}). However, the argument and prefactors are different,
with 
\bmath \label{eq:tau_asym=}
\beqa
 \tau = K(k_1)/(y_1-y_3), \; {\rm for} \; \disc > 0, \\
 \tau = K(k_2)/\sqrt{H_2}, {\rm for} \; \disc < 0,\\
 \tau = \infty, {\rm for} \; \disc = 0 , g_3 \leq 0.
\eeqa
\emath
For $\disc >0, k_1^2=(y_2-y_3)/(y_1-y_3)$ where the roots $y_i$ of
$h(y)$ are all real, $y_i = 
-\sqrt{g_2/3}\cos([\theta+2\pi(i-1)]/3)$ and 
$\theta = \arccos(\sqrt{27g_3^2/g_2^3}).$
For $\disc=-|\disc|<0,k_2=1/2-3 y_2/4(3y_2^2-g_2)$ where 
$y_2$ is the only real root,
$y_2 = [(g_3+\sqrt{-\disc/27})^{1/3}+ (g_3-\sqrt{-\disc/27})^{1/3}]/2$.
Thus the inverse oscillation period $1/\tau$, for $\disc\neq0$, is obtained
as above
in terms of $\Lambda$ and $\Delta E$, with $1/\tau=0$ at 
$\Lambda=\Lambda_c(\Delta E)$.


\begin{thebibliography}{10}

\bibitem{bose-einstein}
{S.~N.~Bose, Z. Phys. {\bf 26}, 178 (1924); A.~Einstein, Sitzber. Kgl. Akad.
  Wiss., (1924), p.261; (1925), p.3; For a historical view, see also A. Pais,
  {\em Subtle is the Lord, The Science and the Life of Albert Einstein}
  (Clarendon Press, Oxford, 1982), Ch. 23}.

\bibitem{bec123}
{M.~H.~Anderson, M.~R.~Matthews, C.~E.~Wieman, and E.~A.~Cornell, Science {\bf
  269}, 198 (1995); K.~B.~Davis, M.-O.~Mewes, M.~R.~Andrews, N.~J.~van~Druten,
  D.~S.~Durfee, D.~M.~Kurn, and W.~Ketterle, Phys. Rev. Lett. {\bf 75}, 3969
  (1995); C.~C.~Bradley, C.~A.~Sackett, J.~J.~Tollett, and R.~G.~Hulet, Phys.
  Rev. Lett. {\bf 75}, 1687 (1995)}.

\bibitem{sacimsk}
{D.~M.~Stamper-Kurn {\em et} al., Phys. Rev. Lett. {\bf 80}, 2027 (1998).}

\bibitem{smcisk}
{D.~M.~Stamper-Kurn {\em et} al., cond-mat/9805022}.

\bibitem{amddkk}
{M.~R.~Andrews {\em et} al., Science, {\bf 273}, 84 (1996) }.

\bibitem{binmix98jila2}
{D.~S.~Hall {\em et al}, cond-mat/9805327}.

\bibitem{iasmsk}
{S.~Inouye {\em et} al., Nature {\bf 392}, 15 (1998).}

\bibitem{sollegg}
A.~J. Leggett and F. Sols, Foundations of Physics {\bf 21},  353  (1991).

\bibitem{gpe}
L.~{P}.~{P}itaevskii, {S}ov. {P}hys. {JETP}, {\bf 13}, 451 (1961); E. P. Gross,
  Nuovo Cimento {\bf 20}, 454 (1961); J. Math. Phys. {\bf 4}, 195 (1963).

\bibitem{collect-edwards-stringari}
{M.~Edwards {\em et} al., Phys. Rev. Lett. {\bf 77}, 1671 (1996); S.~Stringari,
  Phys. Rev. Lett. {\bf 77}, 2360 (1996).}

\bibitem{smerfan}
A. Smerzi and S. Fantoni, Phys. Rev. Lett. {\bf 78},  3589  (1997).

\bibitem{kagan_chaos}
Y. Kagan, E.~L. Surkov, and G.~V. Shlyapnikov, Phys. Rev. A {\bf 55},  R18
  (1997).

\bibitem{dodd-rokh-brsfs}
{R.~J.~Dodd {\em et} al. Phys. Rev. A {\bf 56}, 587 (1997); D.~S.~Rokhsar,
  Phys. Rev. Lett. {\bf 79}, 2164 (1997); M.~Benakli, S.~Raghavan, A.~Smerzi,
  S.~Fantoni, and S.~R.~Shenoy, submitted}.

\bibitem{andrews}
M.~R. Andrews, C.~G. Townsend, H.-J. Miesner, D.~S. Durfee, D.~M. Kurn, and W.
  Ketterle, Science {\bf 275},  637  (1997).

\bibitem{binmix98jila}
{D.~S.~Hall, M.~R.~Matthews,J.~R.~Ensher, C.~E.~Wieman, and E.~.~Cornell,
  cond-mat/9804138.}

\bibitem{javan1}
J. Javanainen, Phys. Rev. Lett. {\bf 57},  3164  (1986).

\bibitem{dps}
F. Dalfovo, L. Pitaevskii, and S. Stringari, Phys. Rev. A {\bf 54},  4213
  (1996).

\bibitem{zapsoleg}
I. Zapata, F. Sols, and A. Leggett, Phys. Rev. A {\bf 57},  R28  (1998).

\bibitem{mcww}
C.~J. Milburn, J. Corney, E.~M. Wright, and D.~F. Walls, Phys. Rev. A {\bf 55},
   4318  (1997).

\bibitem{imalewyou}
A. Imamo\mbox{\={g}}lu, M. Lewenstein, and L. You, Phys. Rev. Lett. {\bf 78},
  2511  (1997).

\bibitem{sfgs}
A. Smerzi, S. Fantoni, S. Giovannazzi, and S.~R. Shenoy, Phys. Rev. Lett. {\bf
  79},  4950  (1997).

\bibitem{barone-ohta}
{A.~Barone and G.~Paterno, {\em Physics and Applications of the Josephson
  Effect} (Wiley, New York, 1982); H.~Ohta, in {\em SQUID: Superconducting
  Quantum Devices and their Applications}, edited by H.~D.~Hahlbohm and
  H.~Lubbig (Walter de Gruyter, Berlin, 1977).}

\bibitem{solymar}
L. Solymar, {\em Superconductive tunnelling and applications} (Chapman and
  Hall, London, 1972).

\bibitem{tinkham}
M. Tinkham, {\em Introduction to Superconductivity}, {2$^{\rm nd}$} ed.
  (McGraw-Hill, New York, 1996).

\bibitem{wkw}
{R.~A.~Webb, R.~L. Kleinberg and J.~C. Wheatley, Phys. Lett {\bf 48}A, 421
  (1974); Phys. Rev. Lett {\bf 33}, 145 (1974)}.

\bibitem{leggrmp}
A.~J. Leggett, Rev. Mod. Phys. {\bf 47},  331  (1975).

\bibitem{matsu}
K. Maki and T. Tsuneto, Prog. Theor. Phys. {\bf 52},  773  (1974).

\bibitem{bpsldp}
S. Backhaus, S.~V. Pereverzev, R.~W. Simmonds, A. Loshak, J.~C. Davis, and
  R.~E. Packard, Nature {\bf 392},  687  (1998).

\bibitem{bec1}
M.~H. Anderson, M.~R. Matthews, C.~E. Wieman, and E.~A. Cornell, Science {\bf
  269},  198  (1995).

\bibitem{nota}
{Including gravitational effects, the trap potential of Eq.~(\ref{eq:gpe})
  becomes $V_{trap}({\bf r}) = \frac{1}{2}m \omega_{trap}^2 {\bf r}^2 - mgz =
  \frac{1}{2}m\omega_{trap}^2({\bf r}-{\bf r}_g)^2 - \frac{1}{2}\frac{m
  g^2}{\omega_{trap}^2}$, where the gravitational acceleration $g$, enters only
  as a `sag' or unimportant shift of both the wavefunctions' centers, ${\bf
  r}_g = \left(0,0,\frac{g}{\omega_{trap}^2}\right)$. Gravitational effects are
  relevant in the context of U-tube oscillations in $^4$-He as studied by
  P.~W.~Anderson, Rev. Mod. Phys. {\bf 38}, 298 (1969).}

\bibitem{bbs}
R.~J. Ballagh, K. Burnett, and T.~F. Scott, Phys. Rev. Lett. {\bf 78},  1607
  (1997).

\bibitem{kc}
V.~M. Kenkre and D.~K. Campbell, Phys. Rev. B {\bf 34},  4959  (1986).

\bibitem{fulton}
T.~A. Fulton,  in {\em Superconductor Applications: SQUIDS and Machines},
  edited by B.~B. Schwartz and S. Foner (Plenum, New York, 1976).

\bibitem{gold-halp}
J.~M. Golden and B.~I. Halperin, Phys. Rev. B {\bf 53},  3893  (1996).

\bibitem{plbdp}
S.~V. Pereverzev, A. Loshak, S. Backhaus, J.~C. Davis, and R.~E. Packard,
  Nature {\bf 388},  449  (1997).

\bibitem{bpldp}
S. Backhaus, S.~V. Pereverzev, A. Loshak, J.~C. Davis, and R.~E. Packard,
  Science {\bf 278},  1435  (1998).

\bibitem{pij}
{L.~N.~Bulaevskii {\em et} al., JETP Lett. {\bf 25}, 290 (1977);
  V.~B.~Geshkenbein {\em et} al., Phys. Rev. B {\bf 36}, 25 (1987);
  D.~A.~Wollman {\em et} al., Phys. Rev. Lett. {\bf 74}, 797 (1993).}

\bibitem{rkdbs-note}
{See S.~Raghavan, V.~M.~Kenkre, D.~H.~Dunlap, A.~R.~Bishop, and M.~I.~Salkola,
  Phys. Rev. A {\bf 54}, R1781 (1996), and references therein.}

\bibitem{giovan}
S. Giovanazzi, Ph.D. thesis, SISSA-ISAS, 1998, (unpublished).

\bibitem{samos-ecu}
{V. M. Kenkre in {\em Singular Behavior and Nonlinear Dynamics}, edited by St.
  Pnevmatikos, T. Bountis, and Sp. Pnevmatikos, (World Scientific, Singapore,
  1989)}; V. M. Kenkre, Physica D {\bf 68}, 153 (1993).

\bibitem{scatter}
V.~M. Kenkre and G.~P. Tsironis, Phys. Rev. B {\bf 35},  1473  (1987).

\bibitem{interplay}
S. Raghavan, V.~M. Kenkre, and A.~R. Bishop, Phys. Lett. A {\bf 233},  73
  (1997).

\bibitem{byrd-milne}
{P}.~{F}.~{B}yrd and {M}.~{D}.~{F}riedman, {\em Handbook of elliptic integrals
  for engineers and scientists, $2^{{\rm nd}}$ {\rm ed.}}, {(Springer, Berlin,
  1971).}; {L}.~{M}.~{M}ilne-{T}homson, in {\em Handbook of Mathematical
  Functions}, edited by {M}.~{A}bramowitz and {I}.~{A}.~{S}tegun ({D}over,
  {N}ew {Y}ork, 1970).

\bibitem{tsiphd}
G.~P. Tsironis, Ph.D. thesis, University of Rochester, 1986, (unpublished).

\bibitem{nmer}
J.~D. Andersen and V.~M. Kenkre, Phys. Rev. B {\bf 47},  11134  (1993).

\bibitem{kenkusmodel}
V.~M. Kenkre and M. Ku\mbox{\'{s}}, Phys. Rev. B {\bf 46},  13792  (1992).

\end{thebibliography}

\begin{figure}
\caption{The asymmetric 
double well trap for two Bose-Einstein condensates with
$N_{1,2}$ and $E^0_{1,2}$, the number of particles and the
zero-point energies in the traps 1 and 2 respectively.}
\label{fig:schematic}
\end{figure}
\begin{figure}
\caption{Population imbalance $z(t)$ 
as a function of dimensionless 
time $2 \kstyle t$ (in units of $\hbar$), with 
conditions $\Lambda=10,\phi(0)=0$, symmetric trap. The initial population
imbalance $z(0)$ takes the values: a)0.1, b)0.5, c)0.59, d)0.6, e)0.65. }
\label{fig:sym_evolve:phi00}
\end{figure}
\begin{figure}
\caption{$z(t)$ as a function of $2 \kstyle t$
 with initial conditions $z(0)=0.6,
\phi(0)=\pi$, symmetric trap. $\Lambda$ takes the values: a)0.1, b)1.1,
c)1.111, d)1.2, e)1.25, f)1.3.}
\label{fig:sym_evolve:phi0pi}
\end{figure}
\begin{figure}
\caption{The $\phi$-potential $V(\phi)$ 
(in arb. units) plotted against $\phi/\pi$ for
$\Lambda=0.2,0.4,0.6$. }
\label{fig:phipotential}
\end{figure}
\begin{figure}
\caption{The $z$-potential $W(z)$ 
(in arb. units) plotted against $z$ in a) and c)
and the corresponding $\phi$-evolution shown in b) and d). In a),b)
$\phi(0)=0$, and in c),d) $\phi(0)=\pi$. The values of $z(0)$ are as shown.}
\label{fig:zpotential}
\end{figure}
\begin{figure}
\caption{The inverse period (scaled in units of $2 \kstyle$) 
$1/\tau$ plotted against $z(0)/z_c$ for
a)$\phi(0)=0$, b)$\phi(0)=\pi$. In a), the dashed line corresponds to
$\Lambda=10$,for which the dip occurs at $z_c=0.6$
 and the solid line to $\Lambda=100$, for which $z_c=0.2$. The inset
in a) shows the time-averaged population imbalance $\ave{z}$ as a function
of $z(0)/z_c$. In b), $\Lambda=1.111,z_c=0.6$. }
\label{fig:symperiodave}
\end{figure}
\begin{figure}
\caption{The phase-plane portrait of the dynamical variables $z$ and $\phi$
for $\Lambda/\Lambda_c$ values as marked.For all trajectories, $z(0)=0.6$.
See text for explanation of the markings of the various trajectories.}
\label{fig:phaseplane}
\end{figure}
\begin{figure}
\caption{
Universal curve for data-collapse with
$ (\pi /(\kstyle C \Lambda^2 \tau)$
 (in units of $\hbar$) versus $k^2(\Lambda)$ as in
Eq.~(\ref{ksq=}).} 
\label{fig:symperiod}
\end{figure}
\begin{figure}
\caption{Scaled inverse period $\tau_{ac}/\tau$ plotted against $\Lambda$ for
fixed asymmetric trap parameter $\Delta E=1$, $z(0)=0.1$,
 and $\phi(0)=0,\pi$ initial
values, and $1/\tau_{ac}$ as defined in Eq.~(\ref{eq:tauptauac}). 
 The vertical scale on the left (right)
corresponds to 
$\phi(0)=0(\pi)$. The insets
show time-averaged
 $\langle z \rangle$ against $\Lambda$, for (a), $\phi(0)=0$
and  (b), $\phi(0)=\pi$. }
\label{fig:asym_period}
\end{figure}
\begin{figure}
\caption{Analog of Shapiro effect: dc current
$I_{dc}=\ave{\dot{z}}$
 versus
trap asymmetry parameter scaled in applied frequency,
 $\Delta E/\omega_0$. Here  $z(0)=0.045,
\phi(0)= \pi /2, \Delta E_1 / {\omega_0 \hbar}=3.5$, and 
dashed (thick solid) lines are for $\Lambda = 0 (1000)$.}
\label{fig:shapiro}
\end{figure}

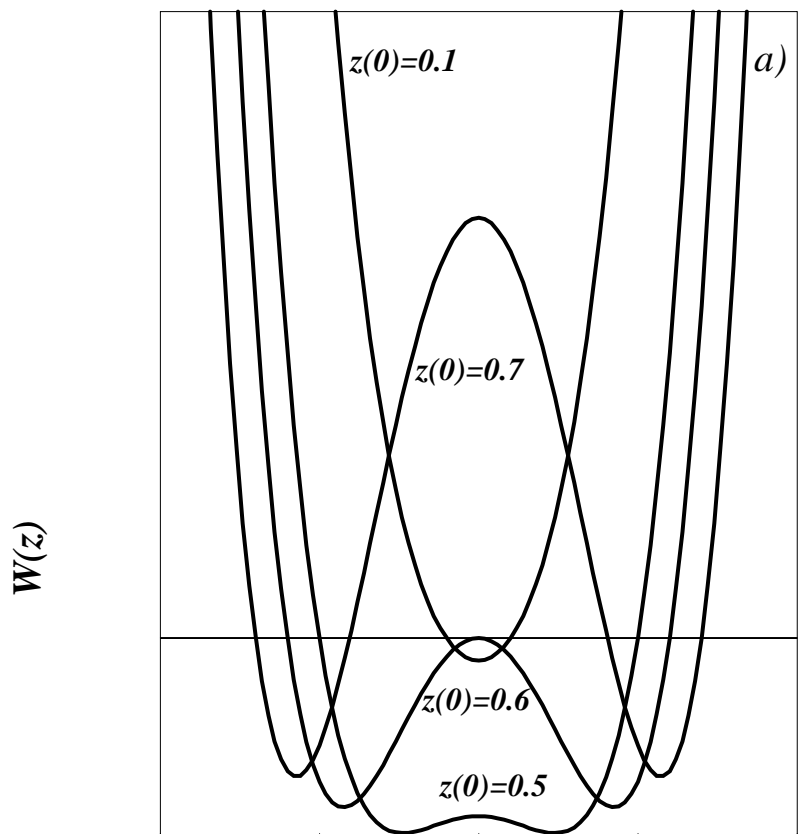
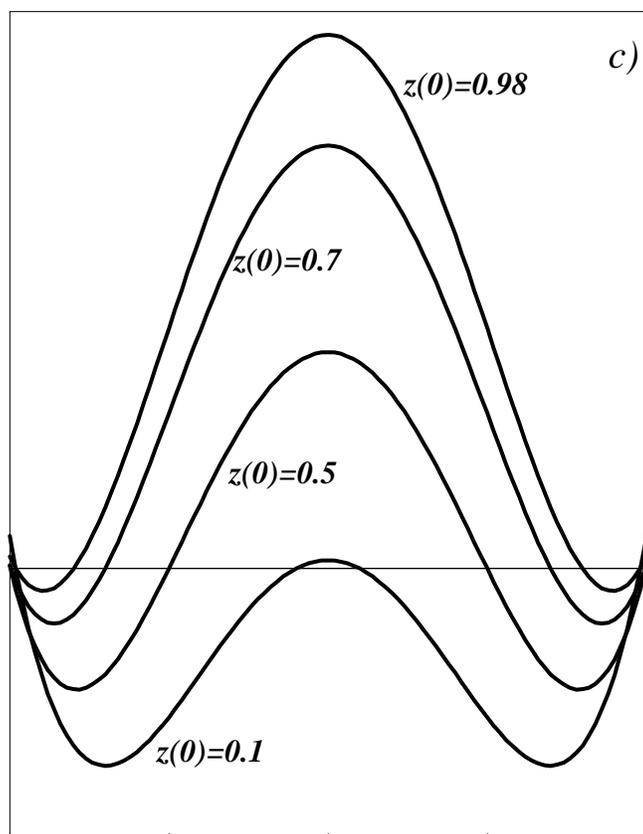
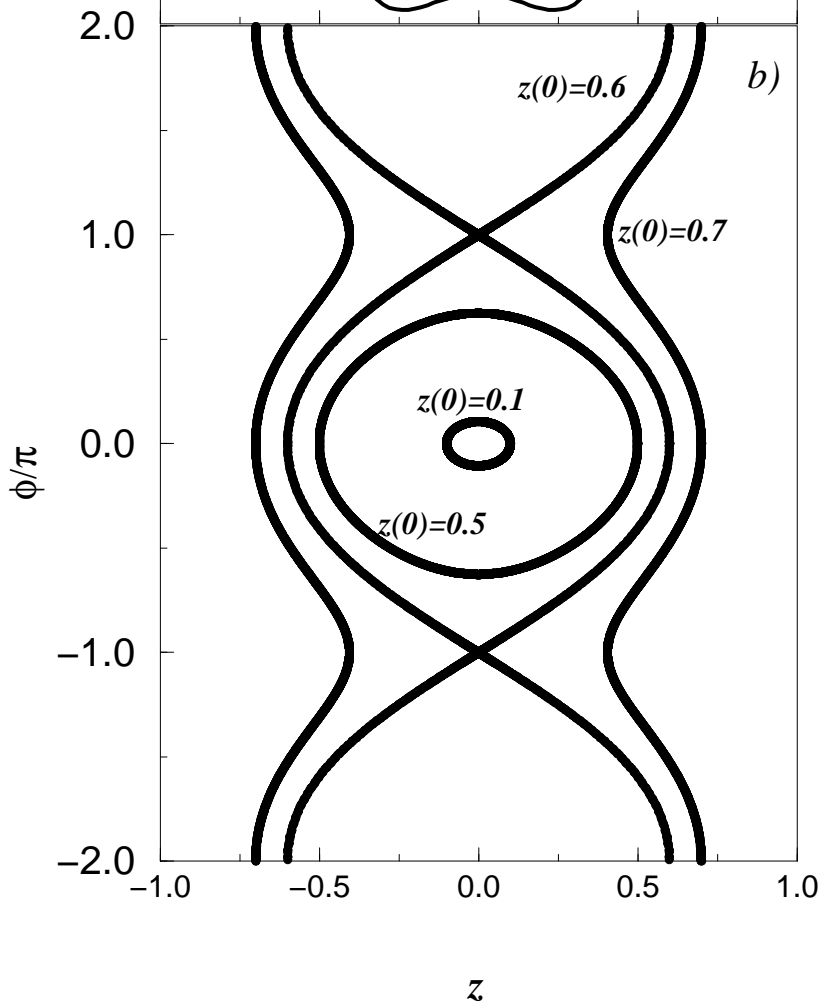
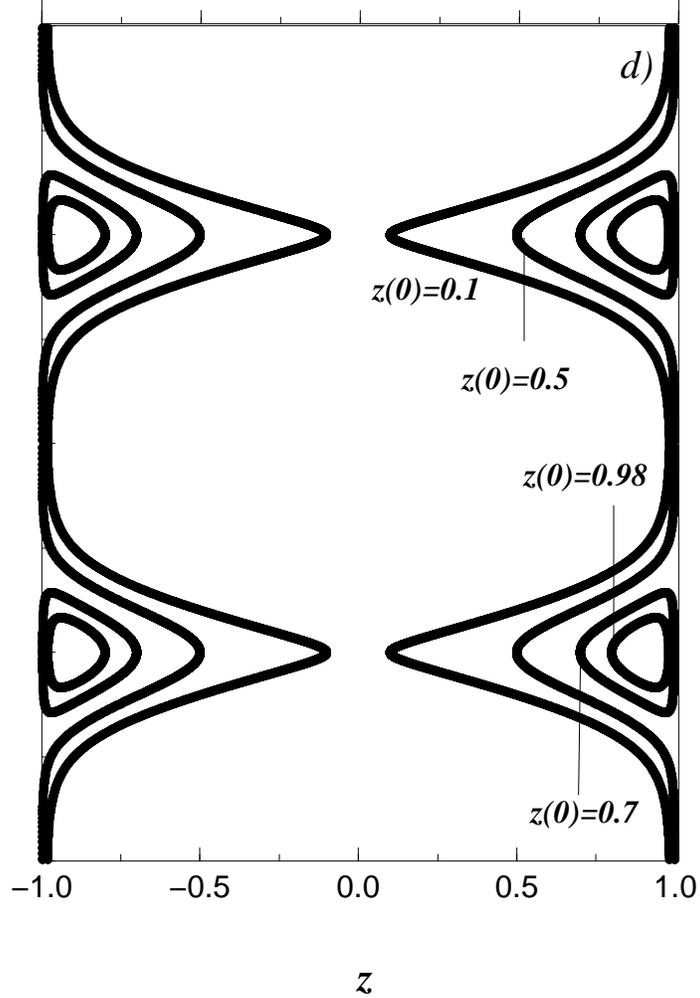
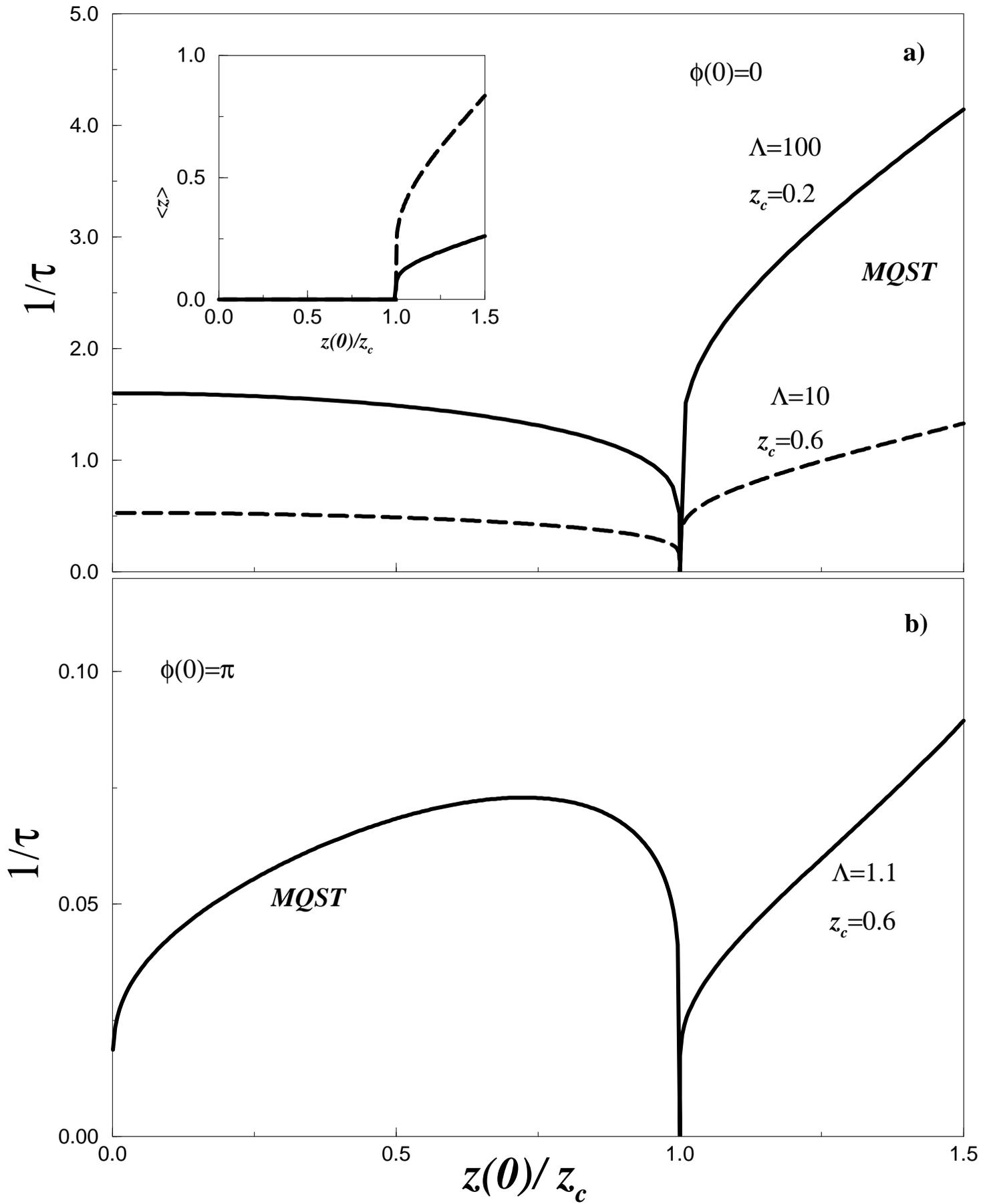
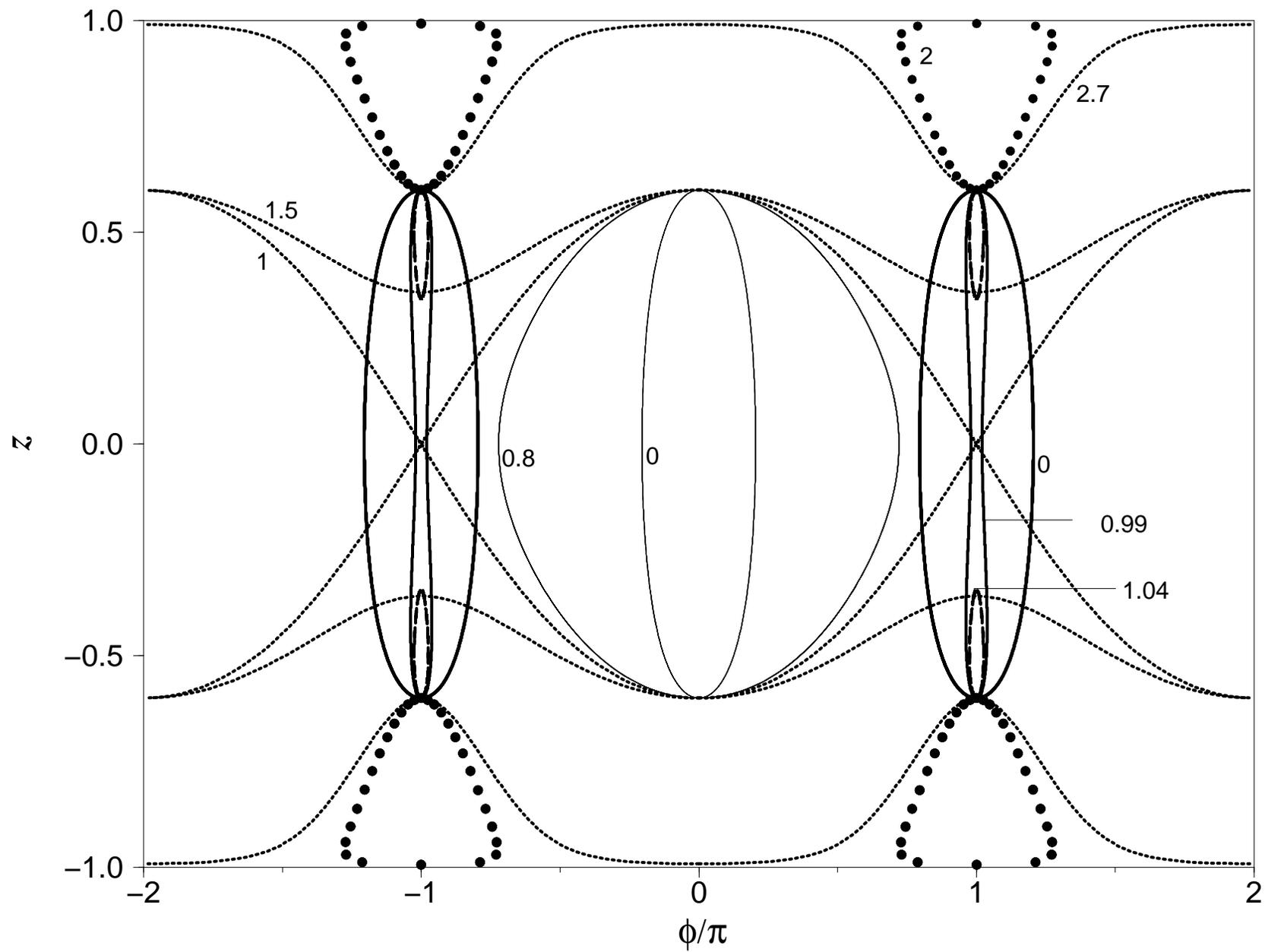
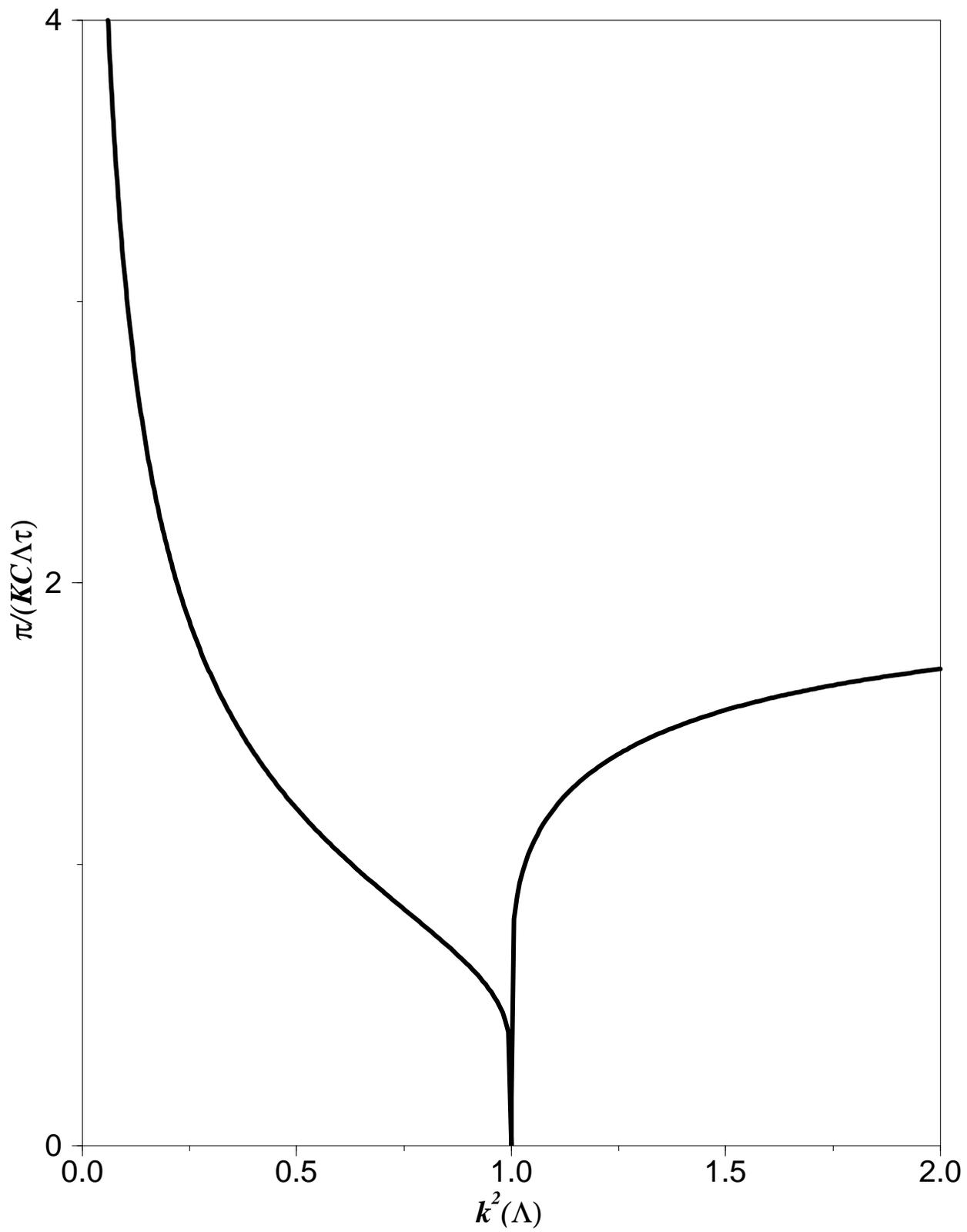
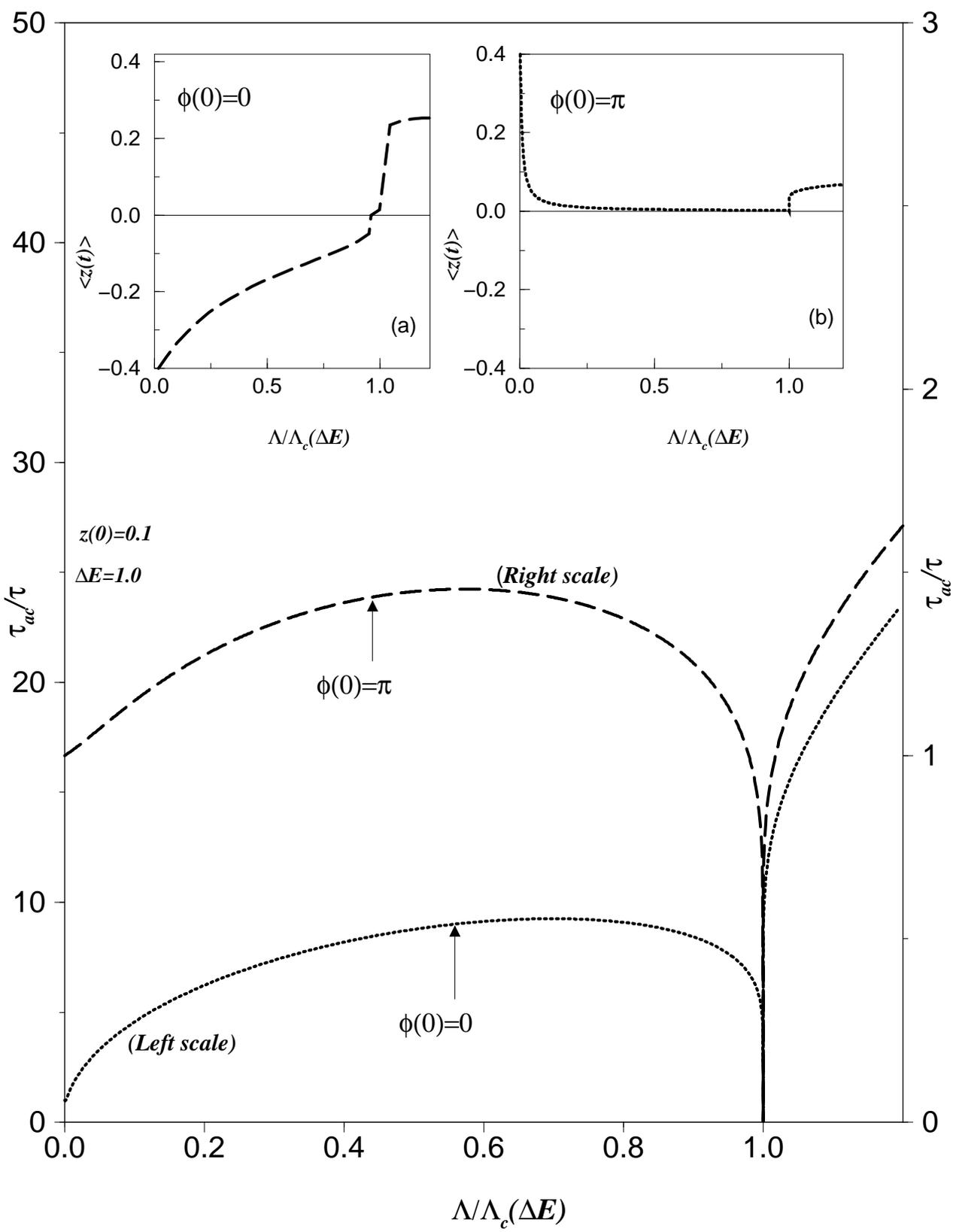
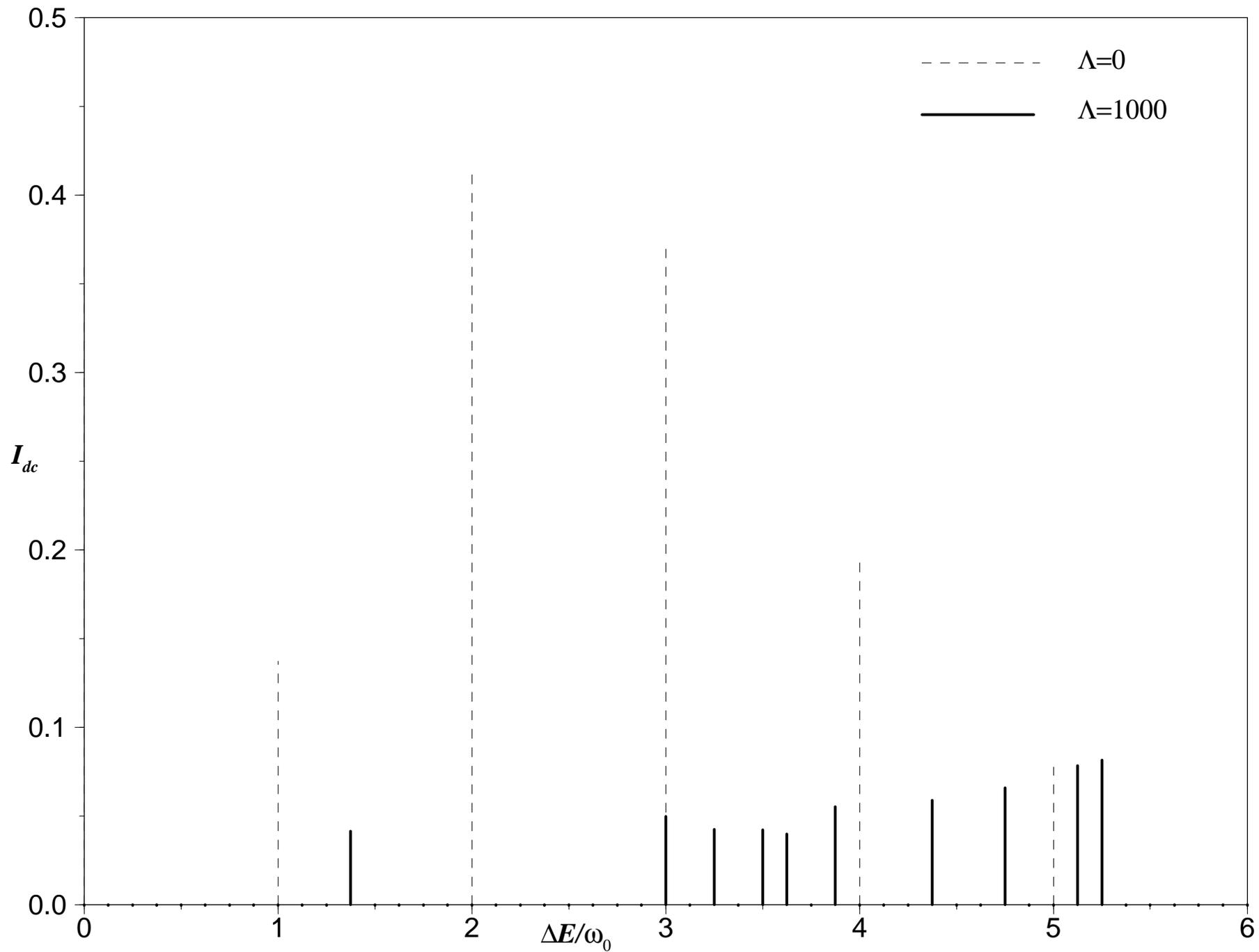
\end{document}